\documentclass[a4paper,twocolumn,10pt]{quantumarticle}
\pdfoutput=1
\usepackage[T1]{fontenc}
\usepackage[utf8]{inputenc}
\usepackage{amsmath}
\usepackage{amssymb}
\usepackage{bm}
\usepackage{graphicx}
\usepackage[export]{adjustbox}
\usepackage{xcolor}
\usepackage{float}
\usepackage{amsthm}
\usepackage[unicode,colorlinks=true]{hyperref}

\usepackage[
  protrusion=true,
  expansion=true,
  stretch=0,
  shrink=0,
  step=1,
  tracking=alltext, 
  letterspace=0
]{microtype}

\usepackage[backend=bibtex]{biblatex}
\theoremstyle{definition}
\newtheorem{definition}{Definition}

\DeclareMathOperator{\diag}{diag}
\DeclareMathOperator{\Tr}{Tr}
\newcommand{\TrA}{\operatorname{Tr}_{\mathrm{A}}}
\newcommand{\TrB}{\operatorname{Tr}_{\mathrm{B}}}
\newcommand{\hatbm}[1]{\hat{\bm{#1}}}

\begin{document}
\title{\texorpdfstring{Quasilinear evolution versus von Neumann selective measurement}{Quasilinear evolution versus von Neumann selective measurement}}
\author{Jakub Rembieli{\'n}ski}
\orcid{0000-0002-5379-4487}
\affiliation{University of Lodz, Faculty of Physics and Applied Informatics, 
Pomorska 149/153, 90-236 Lodz, Poland}
\email{jaremb@uni.lodz.pl}
\author{Karol {\L}awniczak}
\orcid{0000-0002-9530-8846}
\affiliation{University of Lodz, Faculty of Physics and Applied Informatics, 
Pomorska 149/153, 90-236 Lodz, Poland}
\maketitle
\begin{abstract}
In this article, we introduce a new form of quantum selective measurement in which the von Neumann projection postulate is replaced by quasilinear evolution \cite{Rembielinski2020PRR,Rembielinski2021Quantum,Rembielinski2023AnnPhys}, governed by a nonlinear generalization of the von Neumann equation. We demonstrate that this equation preserves the equivalence of quantum ensembles and, consequently, satisfies the no-signalling principle, ensuring consistency with both quantum mechanics and Einstein causality. Our approach eliminates the need for instantaneous, discontinuous state collapse and provides a unified description of the post-measurement quantum state reduction as a form of quantum state evolution. Notably, it does not require invoking concepts such as the quantum state assigned to a classical apparatus. At the same time, the stochastic character of selective measurement and the Born rule remain unchanged.

We present several numerical solutions of the evolution equation for quasilinear selective measurement in two-level quantum systems and compare them with the standard von Neumann projection. The results demonstrate agreement between the two measurement schemes in their fundamental properties.
Furthermore, we investigate phenomena associated with the structural instability of the evolution equation and identify very narrow parameter regions in which the outcomes deviate from those predicted by the von Neumann projection. These regions may offer opportunities to test the proposed approach experimentally.
Finally, using specific analytical solutions, we discuss the Stern--Gerlach experiment within the framework of quasilinear measurement.
\end{abstract}

\section{Introduction}\label{sec:introduction}

Selective measurement, as formulated by John von Neumann, is one of the central elements of quantum theory as it provides an essential means of probing quantum reality. However, quantum mechanics itself does not explain why, during a measurement, the deterministic Schr{\"o}dinger evolution is replaced by a stochastic projection that yields random outcomes in accordance with the Born rule. Moreover, the von Neumann projection is assumed to occur instantaneously, resulting in an abrupt reduction (collapse) of the quantum state. This long-standing issue remains one of the fundamental unresolved problems of quantum mechanics. Enormous theoretical effort has been devoted to addressing what is collectively known as the measurement problem \cite{WheelerZurek1983Book,
Ghirardi1986PRD,
Diosi1987PhysLettA,
Albert1992Book,
Penrose1996GenRelGrav,
dEspagnat2003Book,
Schlosshauer2004RMP,
Ghirardi2005Book,
Maudlin2011Book,
Pusey2012NatPhys,
Bassi2013RMP,
Hance2022JPhysCommun,
Tomaz2025PhilMag}. Nevertheless, despite important achievements like collapse models \cite{Bassi2013RMP} or theory of decoherence \cite{Schlosshauer2004RMP}, to date, there is no consensus among physicists regarding the acceptance of a realistic measurement scheme that either explains or replaces the von Neumann model.

It is evident that unitary evolution cannot transform a mixed state into a pure state. Nonunitary quantum operations, such as Kraus maps \cite{Kraus1983Book}, can have this property; however, in general, the resulting state depends on the initial state. In contrast, the von Neumann projection disregards the initial state: once a given outcome is selected, the post-measurement state is determined solely by the corresponding projector. Above all, however, the von Neumann projection is manifestly a non-linear operation. On the other hand, nonlinear operations---including nonlinear dynamical evolutions---are generally regarded as incompatible with the standard framework of quantum mechanics (see, e.g., \cite{Bielinska2025NJP}). This naturally raises the question of why the nonlinear von Neumann projection is nevertheless accepted as a legitimate quantum operation---an apparent exception within the linear structure of quantum mechanics.

The answer is relatively straightforward: the von Neumann projection preserves the equivalence of quantum ensembles. This observation motivates the question of whether other nonlinear operations (including dynamical evolutions), that also preserve ensemble equivalence, exist, and what role such operations might play in quantum theory. The literature devoted to possible extensions of the quantum-mechanical formalism by the nonlinear one contains many different approaches \cite{BialynickiBirula1976AnnPhys,Weinberg1989TestingQM,Weinberg1989PrecisionQM,Doebner1992PhysLettA,DoebnerGoldin1996PRA,Czachor1998PRA,Mielnik2001PhysLettA} and is still supplemented by new propositions \cite{Kent2005PRA,SergiZloshchastiev2013IJMPB,MeyerWong2014PRA,Helou2017JPCS,Buks2023AQT,Geller2023AdvQuantumTechnol,Buks2024PRA,Ray2025Entropy,Buks2026arXiv}. Nevertheless, as shown by Nicolas Gisin \cite{Gisin1989HelvPhysActa,Gisin1990PhysLettA,Simon2001PRL}, many of them allow superluminal signaling. However, a class of nonlinear transformations, termed quasilinear evolutions, satisfying the requirement of preservation of ensemble equivalence, thus bypassing Gisin's theorem, was found and studied in Refs.~\cite{Rembielinski2020PRR,Rembielinski2021Quantum,Rembielinski2023AnnPhys}.

In this paper, following the approach introduced in \cite{Rembielinski2020PRR,Rembielinski2021Quantum}, we investigate a new form of the deterministic component of selective measurement: a quasilinear evolution defined by a nonlinear generalization of the von Neumann equation. Using the two-level quantum system, we demonstrate that replacing the von Neumann projection with quasilinear evolution reproduces all of its essential properties, including the correct action on entangled states. At the same time, this approach provides a continuous and transparent description of quantum-state dynamics.

An intriguing feature of the proposed framework is the emergence of structural instability, leading to deviations from the standard projection rule within a small region of the parameter space of the evolution equation. We indicate possible experimental consequences of this phenomenon.

Using an analytical solution to the quasilinear evolution equation, we describe the spin state evolution in the context of the famous Stern--Gerlach experiment \cite{GerlachStern1922ZPhysA,GerlachStern1922ZPhysB}.

The paper is organized as follows. In sec.~\ref{sec:ensemble-equivalence}, we discuss the role of quantum statistical ensembles in the context of quantum measurement, emphasizing that the nonlinear von Neumann projection preserves ensemble equivalence. Next, we introduce the notion of quasilinear operations and evolutions, which extends the class of admissible quantum-mechanical maps.

In sec.~\ref{sec:nonlinear-vn}, we study a nontrivial quasilinear evolution equation derived as the simplest nonlinear generalization of the von Neumann equation. Section~\ref{sec:selective-measurement} is devoted to the introduction of quasilinear selective measurement and to a discussion of its fundamental properties. In sec.~\ref{sec:two-level}, we investigate the quasilinear evolution of a two-level quantum system using its Bloch-vector representation and determine the corresponding parameter space of the observable--measurement device pair.

In sec.~\ref{sec:numerical-examples}, we present results of numerical solutions of the quasilinear evolution and compare them with the von Neumann projection. The agreement between the fundamental properties of both types of selective measurement is demonstrated, and the dynamical characteristics of the quasilinear evolution are analyzed. An agreement of the final result of quasilinear evolution with the von Neumann projection is shown.

Section~\ref{sec:critical-angle} addresses phenomena associated with the structural instability of quasilinear evolution in the context of measurement.
Section~\ref{sec:entangled} presents the quasilinear evolution of an entangled state under a local measurement. Unlike generic nonlinear operations, quasilinear evolution acts consistently on product states.
Then, in sec.~\ref{sec:analytic}, we derive a particular analytic solution of the evolution equation. This solution is employed in sec.~\ref{sec:sg} in the description of the Stern--Gerlach experiment. Finally, sec.~\ref{sec:conclusions} contains our conclusions.

\section{Preservation of equivalence of quantum ensembles and the notion of the quasilinear evolution}\label{sec:ensemble-equivalence}

As is well known, ensembles represented by the same density operator cannot be physically distinguished. This holds because the preservation of ensemble equivalence under time evolution is crucial for the no-signaling condition, and thus for excluding violations of Einstein causality. Thus, this issue must be discussed in the context of quantum measurement. From the quantum-mechanical point of view \cite{Paris2012EPJST}, a quantum system is described by statistical ensembles $\{(p_i,\Pi_i)\}$, $i = 1,2,\ldots$, where $p_i$ are probabilities satisfying $\sum_i p_i = 1$. The rank-one projectors $\Pi_i$ (not necessarily mutually orthogonal) represent pure states of the system. An ensemble $\{(p_i,\Pi_i)\}$ is related to the density operator $\rho$ by
\begin{equation}
\rho = \sum_i p_i \Pi_i.
\label{eq:ensemble-density}
\end{equation}
In fact, a given statistical operator $\rho$ represents a variety of ensembles with different probabilities $p_i$ and different projectors $\Pi_i$. Ensembles associated with the same density operator are physically indistinguishable and are therefore regarded as equivalent. Indeed, all physically measurable quantum statistical effects follow from the Born rule, formulated in terms of the density operator and observables. In other words, different preparations that lead to the same density operator define the same quantum state.

Equation~\eqref{eq:ensemble-density} can be generalized to a convex linear combination of density operators,
\begin{equation}
\rho = \sum_{\alpha} \varepsilon_{\alpha} \rho_{\alpha},
\label{eq:convex-density}
\end{equation}
where $\varepsilon_{\alpha} \geq 0$, $\sum_{\alpha} \varepsilon_{\alpha} = 1$, and the density matrices $\rho_{\alpha}$ belong to the convex set $S(\mathcal{H})$ of density operators on the Hilbert space $\mathcal{H}$ of the system. Ensembles of the form $\{(\varepsilon_{\alpha},\rho_{\alpha})\}$ thus generalize the notion of statistical ensembles.

This leads to the following question: Which quantum actions or evolutions preserve the equivalence of ensembles? It is natural to expect that linear actions or evolutions, which preserve the defining properties of density operators, also preserve ensemble equivalence. Consider, for example, a unitary time evolution of the quantum state,
\begin{equation}
\rho(t) = U(t)\rho(0)U^{\dagger}(t).
\label{eq:unitary-evolution}
\end{equation}
Assume that at time $t = t_0 = 0$ the density operator $\rho(0)$ admits two ensemble decompositions,
\begin{equation}
\rho(0) = \sum_{\alpha} \varepsilon_{\alpha} \rho_{\alpha}(0)
       = \sum_{\beta} \widetilde{\varepsilon}_{\beta} \widetilde{\rho}_{\beta}(0).
\label{eq:initial-decompositions}
\end{equation}
These ensembles are therefore equivalent at $t = t_0 = 0$: $\{(\varepsilon_{\alpha},\rho_{\alpha}(0))\} \equiv \{(\widetilde{\varepsilon}_{\beta},\widetilde{\rho}_{\beta}(0))\}$. By the linearity of the unitary action $U(t)$, the evolved state can be written as
\begin{equation}
\rho(t) = \sum_{\alpha} \varepsilon_{\alpha} \rho_{\alpha}(t)
       = \sum_{\beta} \widetilde{\varepsilon}_{\beta} \widetilde{\rho}_{\beta}(t).
\label{eq:evolved-decompositions}
\end{equation}
Hence, the ensembles $\{(\varepsilon_{\alpha},\rho_{\alpha}(t))\}$ and $\{(\widetilde{\varepsilon}_{\beta},\widetilde{\rho}_{\beta}(t))\}$ remain equivalent at time $t$. Strictly speaking, unitary evolution preserves ensemble equivalence. The same conclusion holds for general (possibly non-unitary but linear) Kraus evolution \cite{Kraus1983Book}, as well as for the corresponding equations of motion, such as the von Neumann equation or the Gorini--Kossakowski--Sudarshan--Lindblad (GKSL) master equation \cite{Gorini1976JMP,Lindblad1976CMP}, and for linear quantum operations in general. Note that under linear evolution the coefficients $\varepsilon_{\alpha}$ and $\widetilde{\varepsilon}_{\beta}$ remain unchanged. However, quantum mechanics also involves a nonlinear action, namely, selective measurement. Indeed, the deterministic part of the selective measurement, defined by
\begin{equation}
\rho' = \frac{\Pi \rho \Pi}{\Tr(\Pi \rho)},
\label{eq:projection-map}
\end{equation}
is manifestly nonlinear. Let us apply Eq.~\eqref{eq:projection-map} to the convex combination~\eqref{eq:convex-density}. For simplicity, we consider a two-term decomposition, which does not restrict the generality of the discussion. Let before the projection~\eqref{eq:projection-map} the density matrix $\rho$ be defined by two equivalent ensembles:
\begin{subequations}\label{eq:projection-ensemble}
\begin{align}
\rho &= \varepsilon \rho_a + (1-\varepsilon)\rho_b
     = \widetilde{\varepsilon}\,\widetilde{\rho}_a + (1-\widetilde{\varepsilon})\widetilde{\rho}_b,
\label{eq:projection-ensemble-before}
\end{align}
with $0 \leq \varepsilon \leq 1$ and $0 \leq \widetilde{\varepsilon} \leq 1$. After the projection~\eqref{eq:projection-map}, they are mapped into two equivalent ensembles,
\begin{align}
\rho' &= \varepsilon' \rho'_a + (1-\varepsilon')\rho'_b
      = \widetilde{\varepsilon}'\,\widetilde{\rho}'_a + (1-\widetilde{\varepsilon}')\widetilde{\rho}'_b,
\label{eq:projection-ensemble-after}
\end{align}
\end{subequations}
where $\varepsilon' = \frac{\varepsilon\,\Tr(\Pi \rho_a)}{\Tr(\Pi \rho)}$ and $\widetilde{\varepsilon}' = \frac{\widetilde{\varepsilon}\,\Tr(\Pi \widetilde{\rho}_a)}{\Tr(\Pi \rho)}$, so $0 \leq \varepsilon' \leq 1$ and $0 \leq \widetilde{\varepsilon}' \leq 1$.

We therefore observe that convex combinations are mapped to convex combinations, and that equivalent ensembles are transformed into equivalent ones. However, in contrast to the linear case, the coefficients appearing in the convex decompositions are modified by the nonlinear action of selective measurement.

As shown by Gisin \cite{Gisin1989HelvPhysActa,Gisin1990PhysLettA}, general nonlinear evolutions violate the no-signaling principle by destroying ensemble equivalence. The argument leading to this conclusion, however, does not apply to the nonlinear quantum operations associated with selective measurement, as demonstrated above. Consequently, one may expect the existence of a class of nonlinear quantum actions and evolutions that preserve ensemble equivalence and are therefore admissible within quantum mechanics. Indeed, recent works \cite{Rembielinski2020PRR,Rembielinski2021Quantum,Rembielinski2023AnnPhys} have introduced the notion of a \emph{convex quasilinear map}, which extends the linear structure of quantum mechanics to a class of nonlinear transformations that preserve the equivalence of quantum ensembles.

\begin{definition}[of quasilinear map]
A map $\Phi:S\rightarrow S$ of the manifold of density matrices $S(\mathcal{H})$ is convex quasilinear if for all $\rho_i \in S$ and $\varepsilon_i \in \langle 0,1\rangle$ satisfying $\sum_i \varepsilon_i = 1$, there exists a set $\varepsilon'_k \in \langle 0,1\rangle$, $\sum_k \varepsilon'_k = 1$, such that
\begin{equation}
\Phi\left(\sum_i \varepsilon_i \rho_i\right) = \sum_k \varepsilon'_k \Phi(\rho_k).
\label{eq:quasilinear-map}
\end{equation}
\end{definition}

Evidently, the projection map~\eqref{eq:projection-map}, $\Phi(\rho) = \frac{\Pi \rho \Pi}{\Tr(\Pi \rho)}$, belongs to this class.

The notion of convex quasilinearity can be extended to deterministic time evolutions realized as a one-parameter semigroup evolution $\Phi_t$ of the convex set of density operators $S$ into $S$, i.e.\ $\Phi_t(\rho_0) = \rho(t)$ for $\rho_0 = \rho(0) \in S$ and $\Phi_{\tau}\left(\Phi_t(\rho_0)\right) = \Phi_{t+\tau}(\rho_0) = \rho(t+\tau)$.

\begin{definition}[of quasilinear evolution]
An evolution $\Phi_t$ is convex quasilinear if for all density operators from $S$, $\Phi_t$ satisfies the following condition:
\begin{equation}
\Phi_t\left(\varepsilon_0 \rho_{a0} + (1-\varepsilon_0)\rho_{b0}\right)
=
\varepsilon(t)\Phi_t(\rho_{a0}) + (1-\varepsilon(t))\Phi_t(\rho_{b0}),
\label{eq:quasilinear-evolution}
\end{equation}
provided that $0 \leq \varepsilon(t) \leq 1$ for each $t$.
\end{definition}

This means that quasilinear evolution $\Phi_t$ transforms a convex combination $\varepsilon_0 \rho_{a0} + (1-\varepsilon_0)\rho_{b0}$, given at time $t_0 = 0$, into the convex combination $\varepsilon(t)\rho_a(t) + (1-\varepsilon(t))\rho_b(t)$ at time $t$. Parenthetically, linear evolutions form a subfamily in this class.

\section{Nonlinear generalization of the von Neumann equation}\label{sec:nonlinear-vn}

In contrast to general nonlinear evolutions, a convex quasilinear map preserves---by definition---the convex structure of quantum ensembles as well as ensemble equivalence. A comprehensive analysis and illustrative examples of quasilinear evolutions are presented in Ref.~\cite{Rembielinski2021Quantum}, where the following quasilinear generalization of the GKSL master equation was derived:
\begin{equation}
\begin{split}
\hbar \frac{d\rho}{dt}
&=
-i[H,\rho]
+ \{G,\rho\} +\\
&+ \sum_i L_i \rho L_i^{\dagger}
- \rho\,\Tr\left(\rho\left(2G + \sum_i L_i^{\dagger}L_i\right)\right).
\end{split}
\label{eq:qgksl-general}
\end{equation}
The simplest quasilinear GKSL equation, obtained by elimination of the Lindblad generators $L_i$, is of the form, obtained in \cite{Rembielinski2021Quantum},
\begin{equation}
\hbar \frac{d\rho}{dt} = -i[H,\rho] + \{G,\rho\} - 2\rho\,\Tr(G\rho).
\label{eq:qgksl-simple}
\end{equation}
Here, \emph{H} is the Hamiltonian, while the Hermitian operator \emph{G} generates the nonlinear part of the evolution of the density operator $\rho$. In general, \emph{G} may be chosen to be time dependent. As we see, Eq.~\eqref{eq:qgksl-simple} is a nonlinear generalization of the familiar von Neumann equation. A variant of Eq.~\eqref{eq:qgksl-simple} formulated for pure states was used in Ref.~\cite{Gisin1981JPhysA}  in the specific context of dissipative systems evolution. 

By comparing the traces of the left- and right-hand sides of Eq.~\eqref{eq:qgksl-simple}, we obtain
\begin{equation}
\hbar \frac{d}{dt}\Tr(\rho) = 2(1-\Tr\rho)\,\Tr(G\rho),
\label{eq:trace-balance}
\end{equation}
so the trace condition $\Tr{\rho}=1$ remains satisfied throughout the evolution \eqref{eq:qgksl-simple}, ensuring probability conservation. Moreover, $\rho^2=\rho$ is preserved by the evolution, which means that pure states evolve into pure states. Indeed, to prove this, it is enough to check that the differential identity
\begin{equation}
\frac{d\rho^2}{dt} = \frac{d\rho}{dt}\rho + \rho\frac{d\rho}{dt},
\label{eq:rho-square-derivative}
\end{equation}
with time derivatives of $\rho^2$ and $\rho$ given by Eq.~\eqref{eq:qgksl-simple}
is satisfied. We do it with the help of $\rho G\rho=\rho\Tr{(G\rho)}$ condition
, which follows from the fact that for $\rho^2=\rho$, the operator $\rho$ is the rank-one projector.

Evidently, the energy of a system described by the Hamiltonian \emph{H} is not conserved. Thus, the system is not isolated and undergoes gain and loss processes \cite{Rembielinski2021Quantum,Rembielinski2023AnnPhys}. The operator \emph{G} represents the environment's effective influence on the system's state.

It is straightforward to show that Eq.~\eqref{eq:qgksl-simple} is quasilinear. Indeed, let $\rho$, $\rho_a$, and $\rho_b$ be solutions of this equation. We then substitute the convex combination with time-dependent coefficients
\begin{equation}
\rho(t) = \varepsilon(t)\rho_a(t) + (1-\varepsilon(t))\rho_b(t)
\label{eq:convex-time-combination}
\end{equation}
into the same equation. After elementary calculations, we obtain
\begin{equation}
\frac{2}{\hbar}\Tr\left(G(t)\left(\rho_a(t)-\rho_b(t)\right)\right)
=
\frac{d}{dt}\ln\left(\frac{\varepsilon(t)}{1-\varepsilon(t)}\right).
\label{eq:epsilon-log-derivative}
\end{equation}
Integrating Eq.~\eqref{eq:epsilon-log-derivative} we obtain
\begin{equation}
\begin{split}
&\varepsilon(t)
=\\
&=\frac{
\varepsilon_0 \exp\left(\frac{2}{\hbar}\int_0^t d\tau\,\Tr\left(G(\tau)\left(\rho_a(\tau)-\rho_b(\tau)\right)\right)\right)
}{
1-\varepsilon_0 + \varepsilon_0 \exp\left(\frac{2}{\hbar}\int_0^t d\tau\,\Tr\left(G(\tau)\left(\rho_a(\tau)-\rho_b(\tau)\right)\right)\right)
},
\end{split}
\label{eq:epsilon-solution}
\end{equation}
where $\varepsilon_0=\varepsilon(0)$. Assuming $0 \leq \varepsilon_0 \leq 1$, we see from Eq.~\eqref{eq:epsilon-solution} that $0 \leq \varepsilon(t) \leq 1$. Thus, the convex quasilinearity condition is satisfied. Consequently, Eq.~\eqref{eq:qgksl-simple} satisfies the no-signaling requirement. This means that, similar to the von Neumann projection, the quasilinear Eq.~\eqref{eq:qgksl-simple} defines an acceptable quantum-mechanical operation. Some possible applications of this form of quantum state evolution are given in \cite{Rembielinski2021Quantum} and in \cite{Rembielinski2023AnnPhys}, where the problem of neutrinos moving in the solar plasma was investigated.

\section{Selective measurement as the quasilinear evolution}\label{sec:selective-measurement}

Now, let us consider a scenario where the deterministic part of selective measurement (projection) is replaced by a quasilinear evolution of a quantum elementary system in the form of Eq.~\eqref{eq:qgksl-simple}. The generator $H$ will be reinterpreted as the considered observable, say $\Omega$, satisfying
\begin{equation}
\boxed{\hbar \frac{d\rho}{dt} = -i[\Omega,\rho] + \{G(t),\rho\} - 2\rho\,\Tr\left(G(t)\rho\right),}
\label{eq:measurement-evolution}
\end{equation}
with the initial condition $\rho(0)=\rho_0$. The evolution~\eqref{eq:measurement-evolution}, now interpreted as the deterministic part of selective measurement of a \emph{time-independent} observable $\Omega$, uses an appropriate \emph{time-dependent} generator $G(t)$ encoding device configurations in relation to the quantum system under measurement. Hereafter, we will call the operator $G$ the \emph{driving generator}. It is where the selection of the measurement result, which occurs in accordance with Born's rule, is encoded.

We show that Eq.~\eqref{eq:measurement-evolution} admits a very convenient global solution $\rho(t)$, given by Kraus-like time-dependent operators, namely
\begin{equation}
\boxed{\rho(t) = \frac{K(t)\rho_0 K^{\dagger}(t)}{\Tr\left(K(t)\rho_0 K^{\dagger}(t)\right)},}
\label{eq:kraus-like-solution}
\end{equation}
with the initial condition $K(0)=I$. Notice that here $K(t)$ is defined up to a normalization.

To show the consistent relationship of Eq.~\eqref{eq:kraus-like-solution} with Eq.~\eqref{eq:measurement-evolution}, we calculate $\frac{d\rho(t)}{dt}$ from Eq.~\eqref{eq:kraus-like-solution} and compare it with the right-hand side of Eq.~\eqref{eq:measurement-evolution}. After simple calculations, we conclude that these equations are equivalent if and only if the following relation holds:
\begin{equation}
\boxed{i\hbar \frac{dK(t)}{dt} - \left(\Omega + iG(t)\right)K(t) = 0,
\qquad
K(0)=I.}
\label{eq:k-operator-evolution}
\end{equation}
The nonlinear equation~\eqref{eq:kraus-like-solution} is strikingly similar to the von Neumann projection. It is also quasilinear. This can be shown analogously to the von Neumann projection, see Eqs.~\eqref{eq:projection-map}--\eqref{eq:projection-ensemble-after} and~\eqref{eq:convex-time-combination}, with the time-dependent coefficient $\varepsilon$ of the form
\begin{equation}
\varepsilon(t)
=
\varepsilon_0
\frac{\Tr\left(K(t)\rho_{a0}K^{\dagger}(t)\right)}
{\Tr\left(K(t)\rho_0 K^{\dagger}(t)\right)},
\qquad
\varepsilon(0)=\varepsilon_0.
\label{eq:epsilon-k}
\end{equation}
where $\rho_{a0}=\rho_a(0)$. Obviously, equivalence of $\varepsilon(t)$ given by Eq.~\eqref{eq:epsilon-k} and by Eq.~\eqref{eq:epsilon-solution} holds under the condition~\eqref{eq:k-operator-evolution} only.

We show, in the case of a two-level quantum system, that the evolution~\eqref{eq:measurement-evolution}, or equivalently Eq.~\eqref{eq:kraus-like-solution} with $K(t)$ given by the condition \eqref{eq:k-operator-evolution}, satisfies a number of properties typical for quantum selective measurement realized with the use of rank-one projectors, namely that
\begin{itemize}
\item The final state of the evolution is independent of the initial state.
\item As a result of the measurement, we obtain---with a probability determined by the Born rule---a pure state that is an eigenstate of the observable $\Omega$.
\item The pre-measurement state affects only the probabilities of the possible outcomes, as dictated by the spectral decomposition of the observable and the Born rule.
\item The evolution is well defined on tensor-product states.
\end{itemize}

\section{Quasilinear state evolution in two-level quantum systems}\label{sec:two-level}

In this section, we carefully examine properties of the quasilinear evolution~\eqref{eq:measurement-evolution} in a two-level quantum system, with particular attention paid to phenomena associated with structural instability of the evolution equations. 

In a two-level quantum system, the observable $\Omega$ and its spectral decomposition have the well-known form
\begin{equation}
\Omega(\bm{\omega}) = \frac{1}{2}\,\bm{\omega}\cdot\bm{\sigma}
= \frac{1}{2}\,\omega\sum_{\lambda}\lambda\,\Pi_{\lambda}(\bm{\omega}),
\label{eq:observable-spectrum}
\end{equation}
where $\lambda=\pm1$, $\Pi_{\lambda}(\bm{\omega}) = \frac{1}{2}\left(I+\lambda\,\hatbm{\omega}\cdot\bm{\sigma}\right)$, $\omega = |\bm{\omega}|$, $\hatbm{\omega}=\bm{\omega}/\omega$, and the eigenvalues of $\Omega$ are equal to $\lambda\,\omega/2$. Here $\bm{\sigma}$ is the triplet of Pauli matrices. The density operator $\rho$ is represented by the matrix
\begin{equation}
\rho = \frac{1}{2}\left(I+\bm{n}\cdot\bm{\sigma}\right),
\qquad
\bm{n}^2 \leq 1,
\label{eq:bloch-density}
\end{equation}
where $\bm{n}$ describes the state $\rho$ in the Bloch-vector space. 
Analogously to $\Omega$, the driving operator $G$ is represented by
\begin{equation}
G_\lambda(\bm{g}) = \frac{\lambda}{2}\,\bm{g}(t)\cdot\bm{\sigma},
\label{eq:driving-generator}
\end{equation}
with time-dependent vector $\bm{g}(t)$ for which we also adopt $g(t)=|\bm{g}(t)|$ and $\hatbm{g}=\bm{g}/g$. The generator $G_\lambda(t)$ is understood as an effective generator on the conditional branch with $\lambda=\pm 1$ denoting the outcome selected in a selective measurement. Its probability is given by the Born rule. Once the branch $\lambda$ has been selected, the subsequent evolution is deterministic.

The quasilinear evolution~\eqref{eq:measurement-evolution} of the state $\rho(t)$, generated by the observable $\Omega$ from Eq.~\eqref{eq:observable-spectrum} and the generator $G$ from \eqref{eq:driving-generator} takes the following form:
\begin{equation}
\boxed{\hbar \frac{d\bm{n}(t)}{dt}=\bm{\omega}\times\bm{n}(t)+\lambda\,\bm{g}(t)-\bm{n}(t)\left(\lambda\,\bm{g}(t)\cdot\bm{n}(t)\right),}
\label{eq:bloch-evolution}
\end{equation}
with $\bm{n}(0)\equiv\bm{n}_0$ being the initial-state Bloch vector and $\lambda=\pm1$ corresponding to two trajectories in the space of Bloch vectors. The driving part of Eq.~\eqref{eq:bloch-evolution} selects the trajectory heading to the eigenvectors of the observable $\Omega(\bm{\omega})$ represented by Bloch vectors $\bm{\pi}_{\lambda} = \lambda\,\hatbm{\omega}$. The probability of reaching the final state $\bm{\pi}_\lambda$, and simultaneously the probability of obtaining the corresponding measurement result, based on the initial state $\rho(0)$, is given by Born's rule, i.e.\ $p_{\lambda} = \Tr\left(\rho(0)\Pi_{\lambda}\right) = \frac{1}{2}\left(1+\lambda\,\hatbm{\omega}\cdot\bm{n}_0\right)$. 

Equation~\eqref{eq:bloch-evolution} is form-invariant under the following transformations of the $SL(2\mathbb{C})$ group:
\begin{subequations}\label{eq:sl2c-transform}
\begin{align}
\Omega' + iG' &= A(\Omega+iG)A^{-1},
\label{eq:sl2c-transform-generator}\\
\rho' &= \frac{A\rho A^{\dagger}}{\Tr\left(A\rho A^{\dagger}\right)}.
\label{eq:sl2c-transform-state}
\end{align}
\end{subequations}
where $A \in SL(2\mathbb{C})$. Equation~\eqref{eq:sl2c-transform-generator} defines transformation properties of the pair of vectors $\{\bm{\omega},\bm{g}\}$ belonging to the adjoint representation of $SL(2\mathbb{C})$. Consequently, the Casimir forms $C_1$ and $C_2$, corresponding to the Casimir operators of $SL(2\mathbb{C})$, namely
\begin{subequations}
\begin{align}
C_{1} &= \frac{1}{4}\left(\bm{\omega}^{2}-\bm{g}^{2}\right)
      = \frac{1}{4}\left(\omega^{2}-g^{2}\right),
\label{subeqn:casimirC1}\\
C_{2} &= \frac{1}{2}\,\bm{\omega}\cdot\bm{g}
      = \frac{1}{2}\,\omega\,g\cos\Theta,
\label{subeqn:casimirC2}
\end{align}
\label{eq:casimir-invariants}
\end{subequations}
are invariants under the action of $SL(2\mathbb{C})$. Here $\Theta$ is the angle between $\bm{\omega}$ and $\bm{g}$.

The quasilinear evolution equation~\eqref{eq:bloch-evolution} forms a Riccati-type system of nonlinear equations belonging to the class of structurally unstable differential equations. This implies that, in certain regions of the system's parameter space, small parameter variations may lead to qualitative changes in the global dynamics. We emphasize that structural instability refers to the structure of the dynamical model itself rather than to its initial conditions (see, e.g., \cite{Arnold1994Book}). As shown in \cite{Rembielinski2021Quantum,Rembielinski2023AnnPhys,Kowalski2019AnnPhys}, the system~\eqref{eq:bloch-evolution} exhibits points of structural instability. Therefore, it is reasonable to investigate this issue, particularly in view of its potential experimental consequences. Structurally unstable regions of parameter space are usually associated with critical points of evolution \cite{Heiss2012JPhysA}, in which the corresponding generators become nilpotent. In the present case, the critical region is determined by the zeros of the square of the generator $(\Omega+iG)$,
\begin{equation}
(\Omega+iG)^{2}
=
\left(\frac{1}{2}(\bm{\omega}+i\bm{g})\cdot\bm{\sigma}\right)^2
=
(C_1+iC_2)I,
\label{eq:generator-square}
\end{equation}
Therefore, by means of Eqs.~\eqref{subeqn:casimirC1} and \eqref{subeqn:casimirC2}, the critical region is determined by the two conditions
\begin{subequations}\label{eq:critical-conditions}
\begin{align}
\omega - g &= 0,
\label{eq:critical-conditions-c1}\\
\cos\Theta &= 0,
\label{eq:critical-conditions-c2}
\end{align}
\end{subequations}
the latter implying $\Theta = \pm \pi/2$.

Thus, in a two-level system, we should expect instability of the evolution described by Eq.~\eqref{eq:bloch-evolution} for $g=\omega$ and $\Theta=\pi/2$.

From Eq.~\eqref{eq:generator-square} it follows that
\begin{equation}
C_1+iC_2 = \zeta^2,
\label{eq:eigenvalue-square}
\end{equation}
where $\zeta=\zeta(\bm{\omega},\bm{g})$ is an, in general complex, eigenvalue of the matrix $\Omega(\bm{\omega})+iG(\bm{g})$. Consequently, points of structural instability correspond to zeros of these eigenvalues. In the following, we will discuss the influence of the instability region of the parameter space on the evolution of the Bloch vector $\bm{n}(t)$. To simplify the analysis, we assume that the direction $\hatbm{g}$ of the vector $\bm{g}(t)$ is time-independent, so that only its magnitude $g(t)$ varies in time. Moreover, it is natural to assume that the influence of the generator $G$ on the state of the measured quantum system vanishes outside the space--time region in which the measurement is performed. This implies that $g(t)\to 0$ outside this region. Consequently, the nonnegative function $g(t)$ attains at least one local maximum during the measurement process. In polar coordinates the vectors $\bm{\omega}$, $\bm{g}(t)$ and $\cos\Theta$ take the form
\begin{subequations}\label{eq:polar-parametrization}
\begin{align}
\bm{\omega} &= \omega\,[\sin\alpha\cos\beta,\sin\alpha\sin\beta,\cos\alpha],
\label{eq:polar-parametrization-omega}\\
\bm{g}(t) &= g(t)\,[\sin\theta\cos\varphi,\sin\theta\sin\varphi,\cos\theta],
\label{eq:polar-parametrization-g}\\
\cos\Theta &= \sin\alpha\,\sin\theta\,\cos(\beta-\varphi)+\cos\alpha\,\cos\theta.
\label{eq:polar-parametrization-angle}
\end{align}
\end{subequations}
with $\alpha,\theta,\Theta \in [0,\pi]$ and $\beta,\varphi \in [0,2\pi]$.
In what follows, we examine two model forms of $g(t)$ satisfying these conditions, namely the inverted Morse potential and the Stern--Gerlach potential.

Because of the appearance of critical points for the configuration of $\bm{g}$ perpendicular to a fixed vector $\bm{\omega}$, it is convenient to introduce a coordinate chart parametrized by the angle $\Theta$. This can be done by eliminating in Eq.~\eqref{eq:polar-parametrization-g} the quantities $\sin\varphi$ and $\cos\varphi$ with the use of Eq.~\eqref{eq:polar-parametrization-angle}. As a result we obtain the components of the vector $\bm{g}(t)$ in the form
\begin{equation}
\begin{split}
&g_{1} =
\frac{g(t)}{\sin\alpha}
\Big(
\cos\beta\left(\cos\Theta-\cos\alpha\cos\theta\right)
+\\
&\pm \sin\beta\sqrt{\left(\cos(\alpha-\theta)-\cos\Theta\right)\left(\cos\Theta-\cos(\alpha+\theta)\right)}
\Big),
\\
&g_{2} =
\frac{g(t)}{\sin\alpha}
\Big(
\sin\beta\left(\cos\Theta-\cos\alpha\cos\theta\right)
+\\
&\mp \cos\beta\sqrt{\left(\cos(\alpha-\theta)-\cos\Theta\right)\left(\cos\Theta-\cos(\alpha+\theta)\right)}
\Big),
\\
&g_{3} = g(t)\cos\theta.
\end{split}
\label{eq:g-components}
\end{equation}
with the obvious condition
\begin{equation}
\left(\cos(\alpha-\theta)-\cos\Theta\right)\left(\cos\Theta-\cos(\alpha+\theta)\right)\geq 0.
\label{eq:admissibility-condition}
\end{equation}

\pagebreak

\noindent The inequality~\eqref{eq:admissibility-condition} defines the manifold of admissible configurations of the triples $(\alpha,\theta,\Theta)$ and leads to the simple inequalities for the angles $\alpha,\theta,\Theta$:
\begin{equation}
|\alpha-\theta| \leq \Theta \leq \pi-\left|\pi-(\alpha+\theta)\right|,
\label{eq:angle-constraints}
\end{equation}
while the angle $\beta$ from the interval $[0,2\pi]$ is arbitrary.

\begin{figure}[H]
\centering
\includegraphics[width=0.9167\linewidth]{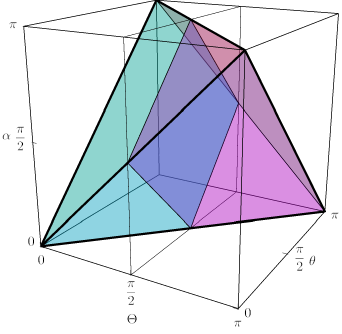}
\caption{The space of parameters $\Theta$, $\theta$, $\alpha$ allowed. The regions corresponding to the measurement outcomes $\lambda$ and $-\lambda$ are coloured teal and purple, respectively. The boundary surface $\Theta = \pi/2$ is highlighted in blue.}
\label{fig:param-space}
\end{figure}

The condition~\eqref{eq:angle-constraints} defines a regular tetrahedron (see Fig.~\ref{fig:param-space}). The points $(\alpha,\theta,\Theta)$ belonging to the tetrahedron, together with the angle $\beta$, determine the relative configurations of the observable $\Omega(\bm{\omega})$ and the generator $G(\bm{g})$. The parameter space depicted in Fig.~\ref{fig:param-space} is divided into two regions corresponding to the respective measurement outcomes. The boundary between them is set by $\Theta=\pi/2$ (compare with Eq.~\eqref{eq:critical-conditions-c2}).

\begin{figure}[H]
\centering
\includegraphics[width=\linewidth]{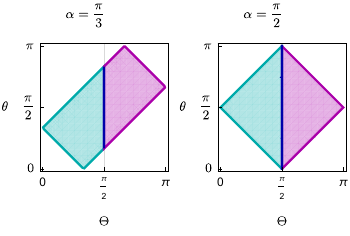}
\caption{Cross-sections of the parameter space from Fig.~\ref{fig:param-space} for selected values of $\alpha$. The areas corresponding to the measurement outcomes $\lambda$ and $-\lambda$ are coloured teal and purple, respectively. The boundary line $\Theta = \pi/2$ is marked in blue.}
\label{fig:param-cross-sections}
\end{figure}

If we fix the observable by selecting a pair of angles $(\alpha,\beta)$, then possible configurations of the generator $G(\bm{g})$ will be determined by pairs $(\theta,\Theta)$ belonging to the cross-section of the tetrahedron defined for a selected value of $\alpha$. In Fig.~\ref{fig:param-cross-sections}, two cross-sections of the tetrahedron for $\alpha=\pi/3$ and $\alpha=\pi/2$ are demonstrated. The division into two regions is clearly visible.

In the next section, we apply the above formalism to investigate the quasilinear evolution describing selective measurement.

\section{Selective measurement as quasilinear evolution in a two-level quantum system}\label{sec:numerical-examples}

In this section, we give representative examples of selective measurement described as deterministic, post-selected quasilinear evolution conditioned on a randomly selected outcome. We consider various initial conditions and various forms of the driving generator $G(\bm{g}(t))$ for a chosen time-independent observable $\Omega(\bm{\omega})$; see Eqs.~\eqref{eq:polar-parametrization-omega}--\eqref{eq:g-components}. We then analyse numerical solutions of the quasilinear evolution equation~\eqref{eq:bloch-evolution} dependent on $\bm{\omega}(\alpha,\beta)$ and $\bm{g}(t,\theta,\Theta)$. We take into account that the measurement is performed in a finite region of space-time. The space-time scale of the considered examples of quasilinear evolutions is encoded in the magnitudes of the vectors $\bm{\omega}$ and $\bm{g}$, i.e.\ in choosing $\omega$ and $g(t)$. This scale is chosen to match that of the Stern--Gerlach experiment, in which both vectors $\bm{\omega}$ and $\bm{g}$ are expressed in terms of the corresponding magnetic fields (see sec.~\ref{sec:sg}). 

\begin{figure}[H]
\centering
\includegraphics[width=\linewidth]{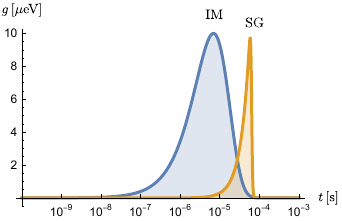}
\caption{The inverted Morse potential $g_{\mathrm{IM}}(t)$ and the Stern--Gerlach potential $g_{\mathrm{SG}}(t)$ for the parameter choice discussed in the text.}
\label{fig:potentials}
\end{figure}

Two forms of the potential $g(t)$ are considered: the inverted Morse potential $g_{\mathrm{IM}}(t)$ and the magnetic potential $g_{\mathrm{SG}}(t)$, appearing in the Stern--Gerlach experiment; see Fig.~\ref{fig:potentials}. The inverted Morse potential $g_{\mathrm{IM}}(t)$ is of the form
\begin{equation}
g_{\mathrm{IM}}(t) = g_0\left(1-\left(1-2e^{-\kappa t}\right)^2\right),
\label{eq:inverse-morse-potential}
\end{equation}
where $g_0$ is a scale factor and $\kappa$ is the shape parameter of the potential. The relevant scales are comparable when, in Eq.~\eqref{eq:inverse-morse-potential}, $\omega \approx g_0 \approx  10^8\,\hbar/\mathrm{s}$. What remains to choose is the shape parameter, which we set to $\kappa = 10^5/\mathrm{s}$.

The second potential considered takes the form
\begin{equation}
g_{\mathrm{SG}}(t)
=
\frac{\mu_B^2\beta^2}{m_{Ag}}t^2\,\mathcal{S}(t),
\label{eq:sg-potential}
\end{equation}
with the smooth switch-off function responsible for the way the magnetic field vanishes at the exit of the experimental device:
\[
\mathcal{S}(t)=\left(
1-\frac{1}{1+\exp\left(-\frac{t-t_{\mathrm{end}}}{t_{w}}\right)}
\right),
\]
where $t_{\mathrm{end}}$ denotes the time at which the measurement ends, while $t_w$ determines the width of the final fall-off of the field. Here the magnetic-field gradient is taken to be $\beta = 10^3\,\mathrm{T/m}$, the mass of the silver atom is $m_{Ag}=1.11\times 10^{-6}\,\mathrm{eV}\,\mathrm{s}^2/\mathrm{m}^2$, and $\mu_B$ denotes the Bohr magneton. A detailed justification of the formula \eqref{eq:sg-potential} is provided in sec.~\ref{sec:sg} devoted to the Stern--Gerlach experiment.

We present both potentials in Fig.~\ref{fig:potentials}.

In the following sections, we mainly use the inverted Morse potential because its flexibility allows us to test how the evolution depends on the form of the driving potential.

\subsection{Quasilinear state evolution for various initial states}\label{subsec:initial-states}

We show that the quasilinear evolution of the state in the presence of a potential $g_{\mathrm{IM}}(t)$ results in a Bloch vector $\bm{\pi}_{\lambda}=\lambda\hatbm{\omega}$ describing an eigenstate of the observable $\Omega(\bm{\omega})$. The following figure contains the examples of the evolution of the Bloch vector $\bm{n}(t)$ from three different initial states: pure, mixed, and fully depolarized. Cases of evolution conditioned on both possible measurement outcomes $\lambda=\pm 1$ are presented side by side.

\begin{widetext}

\begin{figure}[H]
    \centering
    \begin{tabular}{c c c}
        & \makebox[0.4375\textwidth][c]{$\lambda=+1$}
            \makebox[0.4375\textwidth][c]{$\lambda=-1$}
        & \\
        & \includegraphics[valign=c,width=0.875\textwidth]{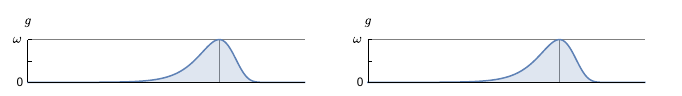} & \\
        \rotatebox[origin=c]{90}{$\bm{n}_0=\left[0,-\frac{1}{\sqrt{2}},-\frac{1}{\sqrt{2}}\right]$} &
        \includegraphics[valign=c,width=0.875\textwidth]{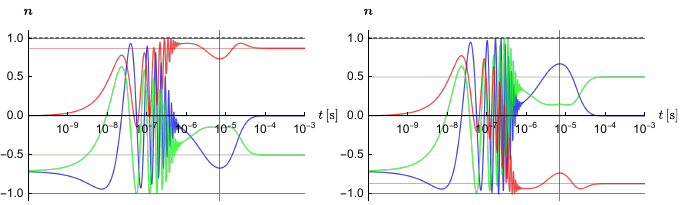} & \\
        \rotatebox[origin=c]{90}{$\bm{n}_0=\frac{1}{\sqrt{2}}\left[\frac{1}{\sqrt{2}},-\frac{1}{\sqrt{2}},0\right]$} &
        \includegraphics[valign=c,width=0.875\textwidth]{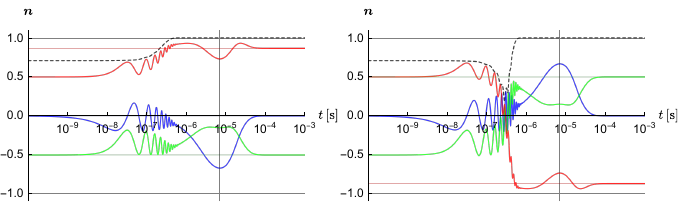} & \\
        \rotatebox[origin=c]{90}{$\bm{n}_0=\bm{0}$} &
        \includegraphics[valign=c,width=0.875\textwidth]{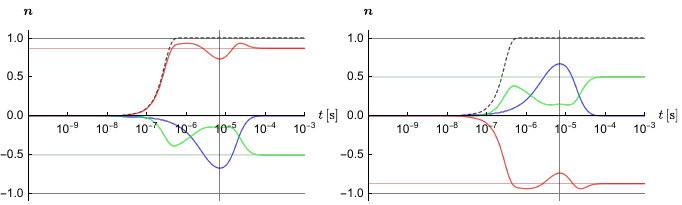} & \\
        & \includegraphics[valign=c,width=0.656\textwidth]{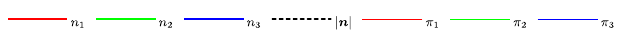} & \\
    \end{tabular}
    \caption{State evolution according to Eq.~\eqref{eq:bloch-evolution} from three different initial states $\bm{n}_{0}$. The vector $\bm{\omega}$ defining the observable $\Omega$ is directed at $\alpha=\pi/2$, $\beta=-\pi/6$, so that it has the form $\bm{\omega}=\omega\left[\frac{\sqrt{3}}{2},-\frac{1}{2},0\right]$, where $\omega = 10^8\,\hbar/\mathrm{s}$. The vector $\bm{g} = g(t)\left[\frac{\sqrt{3}-1}{4},-\frac{\sqrt{3}+1}{4},-\frac{1}{\sqrt{2}}\right]$ is defined by the angles $\theta=3\pi/4$, $\Theta=\pi/3$. Its time-dependent magnitude $g(t)$ is defined as in Fig.~\ref{fig:potentials}, with parameter $\kappa = 10^5/\mathrm{s}$ and peak value $g_0 = 10^8\,\hbar/\mathrm{s}$. The maximum of the potential at $t_{\max}=\ln2/\kappa$ is marked by the vertical line. The two columns show the dynamics conditioned on two possible measurement outcomes, $\lambda=\pm1$. In each plot, the horizontal axis indicates the time $t$ on a logarithmic scale. The red, green, and blue lines denote the components of the state vector $\bm{n}(t)$; the thin straight lines in the same colors denote the respective components of the vector $\bm{\pi}_{\lambda}=\lambda\hatbm{\omega}$ representing the state resulting from the von Neumann projection. The black dashed line is the norm $|\bm{n}(t)|$ of the vector $\bm{n}(t)$. Note that in all cases, the final asymptotic states of evolution coincide with the von Neumann projections. Thus, they are independent of the initial conditions.}
    \label{fig:initial-states}
\end{figure}
\clearpage
\end{widetext}

In all three cases, as the potential $g(t)$ decays to zero, the evolution drives the state toward the eigenstates of the observable $\Omega(\bm{\omega})$. The two possible trajectories corresponding to $\lambda=\pm1$ are realized randomly, with probabilities determined by the Born rule. Importantly, the final asymptotic state is independent of the initial state, exactly as in the von Neumann measurement scheme. These properties hold for all configurations in the parameter space of the model, except in the vicinity of the cross-section defined by the condition $\Theta = \pi/2$ corresponding to the Casimir invariant $C_2=0$; see Eq.~\eqref{eq:critical-conditions-c2} and Fig.~\ref{fig:param-space}. Since the angle $\Theta$ between $\bm{\omega}$ and $\bm{g}$ here is equal to $\pi/3$, which is far from $\pi/2$, $C_2$ is not relevant for the evolution. On the other hand, the detailed dynamics of the evolution depends on both the initial state and the outcome $\lambda$.

In Fig.~\ref{fig:initial-states}, Bloch vectors with non-zero magnitude undergo precession, visible as oscillations of their components. This is reflected in Fig.~\ref{fig:rate-change-initial} as an approximately constant rate of state change. A fully depolarized state does not exhibit such precession. As the driving potential increases, the precession is suppressed, and the state is attracted toward the observable eigenstate. This attraction can even occur with an increase in the rate of change of state. Ultimately, the rate of state change drops to a value close to zero. In the case presented, this happens long before the potential reaches its maximum value, i.e.\ at about $10^{-6}\,\mathrm{s}$. After this time, only a transient and relatively slow deformation of the state takes place in the neighborhood of the maximum of the potential. The critical point of the evolution, associated with $C_1$, occurs when $g(t)=\omega$; see Eq.~\eqref{eq:critical-conditions-c1}. With this particular choice of relative magnitudes $g_0=\omega$, this happens at a single point, namely in the maximum of the potential $t_{\mathrm{crit}}=t_{\mathrm{max}}=\ln2/\kappa\approx 0.69\times10^{-5}\,\mathrm{s}$; see Fig.~\ref{fig:potentials}.  The influence of this critical point on the evolving state is limited to a deformation of the Bloch-vector trajectories $\bm{n}(t)$ and does not affect the final state.

\begin{widetext}

\begin{figure}[H]
    \centering
    \begin{tabular}{c c c}
        & \makebox[0.4375\textwidth][c]{$\lambda=+1$}
            \makebox[0.4375\textwidth][c]{$\lambda=-1$}
        & \\
        & \includegraphics[valign=c,width=0.875\textwidth]{general_g.pdf} & \\
        \rotatebox[origin=c]{90}{$\bm{n}_0=\left[0,-\frac{1}{\sqrt{2}},-\frac{1}{\sqrt{2}}\right]$} &
        \includegraphics[valign=c,width=0.875\textwidth]{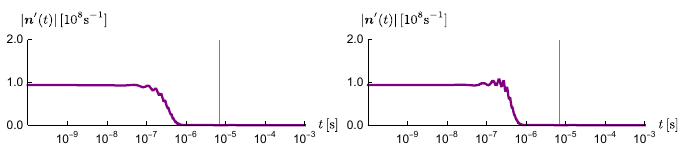} & \\
        \rotatebox[origin=c]{90}{$\bm{n}_0=\frac{1}{\sqrt{2}}\left[\frac{1}{\sqrt{2}},-\frac{1}{\sqrt{2}},0\right]$} &
        \includegraphics[valign=c,width=0.875\textwidth]{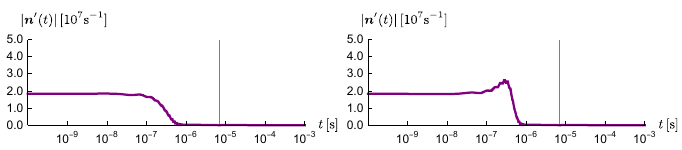} & \\
        \rotatebox[origin=c]{90}{$\bm{n}_0=\bm{0}$} &
        \includegraphics[valign=c,width=0.875\textwidth]{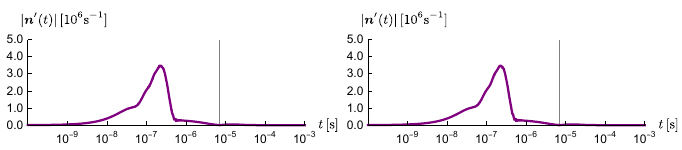} & \\
    \end{tabular}
    \caption{Rate of change of the Bloch vector under quasilinear evolution starting from three different initial states $\bm{n}_{0}$ from Fig.~\ref{fig:initial-states}.}
    \label{fig:rate-change-initial}
\end{figure}
\clearpage
\end{widetext}

In summary, the final state of the quasilinear evolution coincides with the corresponding von Neumann projection and does not depend on the initial state.

\subsection{Independence of the final state from variations of the driving potential}\label{subsec:potential-variations}

In practice, the experimental determination of the spectrum and eigenstates of an observable can be carried out using different measurement devices. Within our model, this implies that for a fixed observable $\Omega(\bm{\omega})$, selected outcome $\lambda$, the final state should remain unchanged for a broad class of variations of the driving generator $G(\bm{g}(t,\theta,\Theta))$; that is, the final state must be an eigenstate of the observable $\Omega$ corresponding to $\lambda$. We demonstrate that this condition is satisfied under changes in the scale, orientation, and functional form of $\bm{g}(t)$.

\vspace{1ex}

\subsubsection{Variation of the scale \texorpdfstring{$g_{0}$}{g0}}

The potential $\bm{g}(t)$ may have a different scale, quantified by $g_0$. The following figure presents the state dynamics under the scale $g_0$ much lower and much higher than in previous examples. The remaining parameters of the evolution are the same as in Fig.~\ref{fig:initial-states}. The initial state is taken to be one more mixed state $\bm{n}_0=\left[0,-\frac12,-\frac12\right]$.

\begin{widetext}

\begin{figure}[H]
    \centering
    \begin{tabular}{c c c}
        & \makebox[0.4375\textwidth][c]{$\lambda=+1$}
            \makebox[0.4375\textwidth][c]{$\lambda=-1$}
        & \\
        & \includegraphics[valign=c,width=0.875\textwidth]{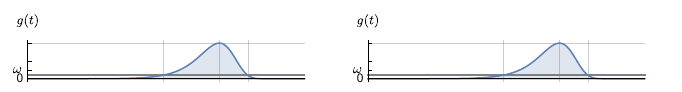} & \\
        \rotatebox[origin=c]{90}{$g_0=10\omega$} &
        \includegraphics[valign=c,width=0.875\textwidth]{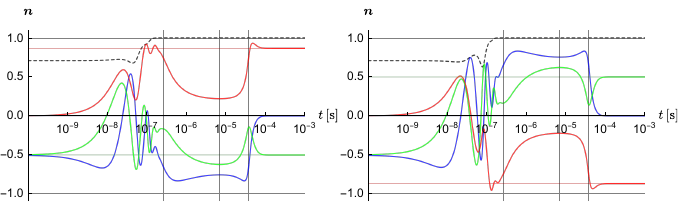} & \\
        & \includegraphics[valign=c,width=0.875\textwidth]{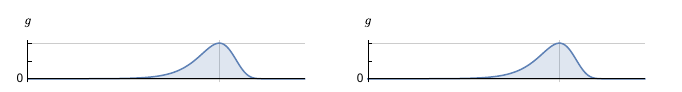} & \\
        \rotatebox[origin=c]{90}{$g_0=\omega/100$} &
        \includegraphics[valign=c,width=0.875\textwidth]{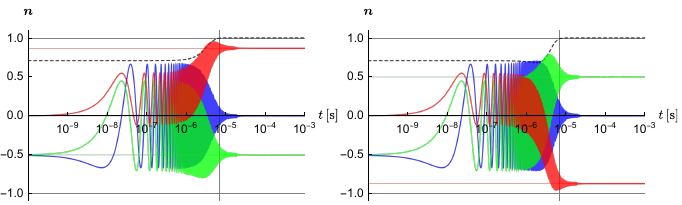} & \\
        & \includegraphics[valign=c,width=0.656\textwidth]{general_legend.pdf} & \\
    \end{tabular}
    \caption{State dynamics for the scale $g_{0}$ of the $g$-term set much higher (upper graph) and much lower (lower graph) than in previous examples. The vector $\bm{\omega}$, the direction $\hat{\bm{g}}$, and the form of $g(t)$ are kept the same as in Fig.~\ref{fig:initial-states}. Initial state $\bm{n}_{0} = [0,-\frac{1}{2},-\frac{1}{2}]$.}
    \label{fig:scale-variation}
\end{figure}
\clearpage
\end{widetext}

The time required to reach the final state depends on the magnitude of the potential $g(t)$, quantified by $g_0$. As this magnitude decreases, the time needed to approach the asymptote increases. For low values of $g_0 \ll \omega$ (upper graph), no noticeable disturbance of the state is observed after the precession has ceased, in contrast to the behavior seen in previous cases. In general, the amplitude of such disturbances increases with $g_0$. When $g_0$ exceeds $\omega$ (lower graph), the characteristic disturbance extends over the entire region between the points where the Casimir form $C_1=0$, i.e.\ $g(t_{\pm})=\omega$; see Eq.~\eqref{eq:critical-conditions-c1}. Here $t_{-}$ and $t_{+}$ denote the intersection times of the potential $g(t)$ with the constant $\omega$. The times $t_{\pm}$ are given by
\[
t_{\pm}
=
\frac{1}{\kappa}
\ln\left(
2\frac{g_0}{\omega}
\left(
1 \pm \sqrt{1-\frac{\omega}{g_0}}
\right)
\right).
\] 
Although the scale variation and the critical phenomena arising from the Casimir form $C_1$ influence the internal dynamics of the evolution, they do not affect the final state, which coincides with the von Neumann projection result. However, below a certain threshold value of $g_0$, the potential becomes insufficient to suppress oscillations. This regime will be discussed in the following sections.

\subsubsection{Variation of the orientation of \texorpdfstring{$\bm{g}$}{g}}

The second thing we can change in the potential $\bm{g}(t)$ is its orientation. Figure~\ref{fig:orientation-variation} presents the state dynamics for two orientations of $\bm{g}$. We focus on the  angle $\Theta$ determining the orientation of $\bm{g}$ relative to the vector $\bm{\omega}$. The initial state is $\bm{n}_0=\left[0,-\frac12,-\frac12\right]$ again. The remaining parameters of the evolution are the same as in Fig.~\ref{fig:initial-states}.

\begin{widetext}

\begin{figure}[H]
    \centering
    \begin{tabular}{c c c}
        & \makebox[0.4375\textwidth][c]{$\lambda=+1$}
            \makebox[0.4375\textwidth][c]{$\lambda=-1$}
        & \\
        & \includegraphics[valign=c,width=0.875\textwidth]{general_g.pdf} & \\
        \rotatebox[origin=c]{90}{$\Theta=0, \theta=\frac{\pi}{2}$} &
        \includegraphics[valign=c,width=0.875\textwidth]{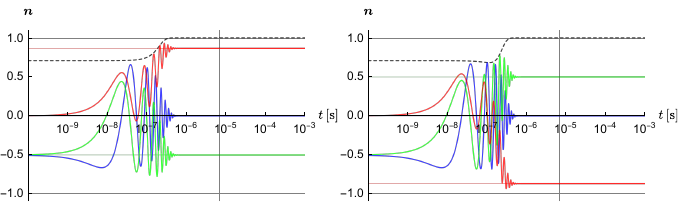} & \\
        \rotatebox[origin=c]{90}{$\Theta=\frac{\pi}{3}, \theta=\frac{\pi}{6}$} & \includegraphics[valign=c,width=0.875\textwidth]{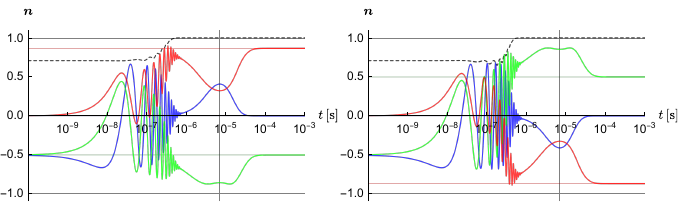} & \\
        & \includegraphics[valign=c,width=0.656\textwidth]{general_legend.pdf} & \\
    \end{tabular}
    \caption{State dynamics for two substantially different angles $\Theta$ between $\bm{g}$ and $\bm{\omega}$. The vector $\bm{\omega}$ and the function $g(t)$ are kept the same as in Fig.~\ref{fig:initial-states}; however, the orientation of $\bm{g}$ is different. The initial state is $\bm{n}_{0} = [0,-\frac{1}{2},-\frac{1}{2}]$.}
    \label{fig:orientation-variation}
\end{figure}
\clearpage
\end{widetext}

As we observe, the orientation of the vector $\bm{g}$ affects the internal dynamics of the evolution, specifically
the shape of the disturbance observed after the precession has ceased. The scale of the disturbance also depends on the angle $\Theta$. Note that for $\Theta=0$, this disturbance does not occur at all. For $\Theta=\pi/3$, disturbances related to the maximum of the potential are observed near $t_{\mathrm{max}}=\ln(2)/\kappa$. The behavior of the state in the vicinity of $\Theta=\pi/2$ and for $\Theta>\pi/2$ is the subject of sec.~\ref{sec:critical-angle}.

\subsubsection{Variation of the form of \texorpdfstring{$g(t)$}{g(t)}}

Finally, the functional form of the potential $g(t)$ can be chosen in various ways. In the case of the inverted Morse potential, the shape parameter $\kappa$ may be adjusted. More generally, the inverted Morse potential can be replaced by a completely different potential, provided that it decays to zero---at least asymptotically---outside the measurement region. In the upper part of the Fig.~\ref{fig:form-variation}, we present the state dynamics driven by the inverted Morse potential with the parameter $\kappa = 10^{7}/\mathrm{s}$, which shifts the maximum of the potential toward the origin. In the lower part of Fig.~\ref{fig:form-variation}, we present the solution to the evolution equation~\eqref{eq:bloch-evolution} obtained using the Stern--Gerlach potential; see Eq.~\eqref{eq:sg-potential}.

\begin{widetext}

\begin{figure}[H]
    \centering
    \begin{tabular}{c c c}
        & \makebox[0.4375\textwidth][c]{$\lambda=+1$}
            \makebox[0.4375\textwidth][c]{$\lambda=-1$}
        & \\
        & \includegraphics[valign=c,width=0.875\textwidth]{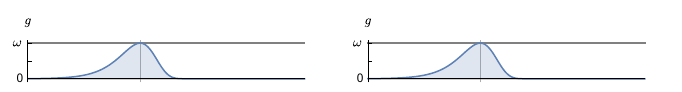} & \\
        \rotatebox[origin=c]{90}{IM, $\kappa=10^7 /\mathrm{s}$} &
        \includegraphics[valign=c,width=0.875\textwidth]{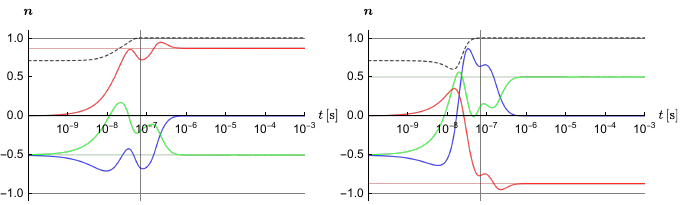} & \\
        & \includegraphics[valign=c,width=0.875\textwidth]{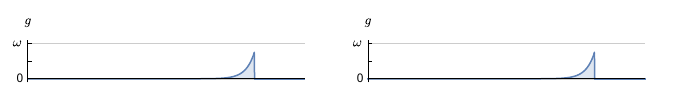} & \\
        \rotatebox[origin=c]{90}{SG} &
        \includegraphics[valign=c,width=0.875\textwidth]{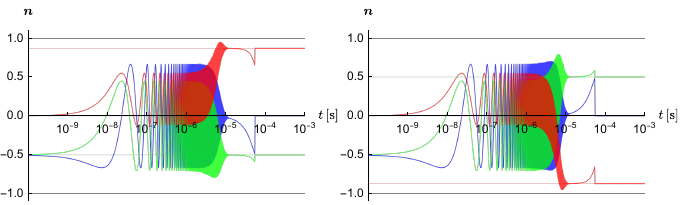} & \\
        & \includegraphics[valign=c,width=0.656\textwidth]{general_legend.pdf} & \\
    \end{tabular}
    \caption{State dynamics for two different forms of the function $g(t)$: the inverted Morse with the parameter $\kappa$ much higher than in the previous examples, and the potential of the Stern--Gerlach experiment. The vector $\bm{\omega}$ and the direction of $\bm{g}$ are kept the same as in Fig.~\ref{fig:initial-states}. The initial state is $\bm{n}_{0} = [0,-\frac{1}{2},-\frac{1}{2}]$.}
    \label{fig:form-variation}
\end{figure}
\clearpage
\end{widetext}

For the inverted Morse potential with $\kappa=10^7/\mathrm{s}$, we observe that the time required to approach the asymptotic states is shorter than in Fig.~\ref{fig:initial-states}. The Stern--Gerlach potential leads to dynamics different from those observed for the inverted Morse potential. In particular, the time required to approach the asymptotic states is considerably longer than in the former case. Nevertheless, the final asymptotic states in both cases are identical to those shown in Fig.~\ref{fig:initial-states}. Thus, the final state is insensitive to the specific form of the potential, provided that it satisfies reasonable measurement conditions.

Taking into account these three aspects of the final state's robustness with respect to changes in the potential,
we can state that the final asymptotic state does not depend on the details of the driving potential and coincides with the state obtained by the von Neumann projection. However, there exists an exception: the vicinity of the critical point.

\section{Quasilinear measurement in the vicinity of the critical angle \texorpdfstring{$\Theta=\pi/2$}{Theta = pi/2}}\label{sec:critical-angle}

From sec.~\ref{sec:numerical-examples} we know that the critical regions of the parameter space of the model corresponding to the vanishing of the Casimir invariant $C_1$ do not affect the final state of evolution. A different situation arises in the case when the second Casimir invariant takes zero value, which holds for $\bm{g}\perp\bm{\omega}$, or when $g_0\ll\omega$.

\subsection{In the vicinity of the critical angle \texorpdfstring{$\Theta=\pi/2$}{Theta = pi/2}}

Figure~\ref{fig:both-sides-border} presents what happens when we cross this line. The evolution of the state takes on a particularly interesting form when, during evolution, the value of $g(t)$ exceeds $\omega$, which means $C_1$ passes through $0$ twice.

\begin{widetext}

\begin{figure}[H]
    \centering
    \begin{tabular}{c c c}
        & \makebox[0.4375\textwidth][c]{$\lambda=+1$}
            \makebox[0.4375\textwidth][c]{$\lambda=-1$}
        & \\
        & \includegraphics[valign=c,width=0.875\textwidth]{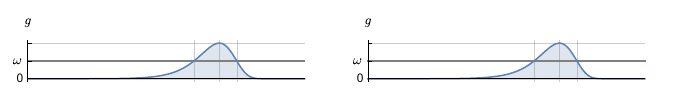} & \\
        \rotatebox[origin=c]{90}{$\Theta=\frac{\pi}{2}-0.01$} &
        \includegraphics[valign=c,width=0.875\textwidth]{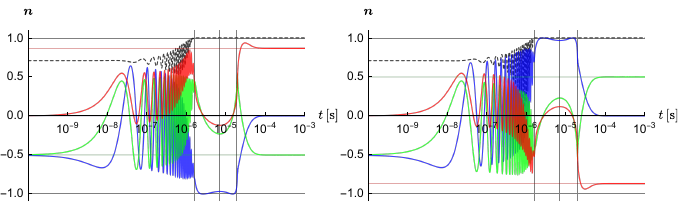} & \\
        \rotatebox[origin=c]{90}{$\Theta=\frac{\pi}{2}+0.01$} &
        \includegraphics[valign=c,width=0.875\textwidth]{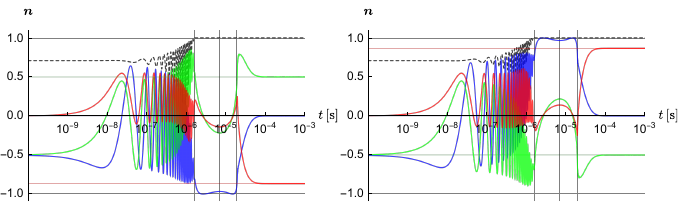} & \\
        & \includegraphics[valign=c,width=0.656\textwidth]{general_legend.pdf} & \\
    \end{tabular}
    \caption{State dynamics on both sides of the $\Theta = \pi/2$ borderline ($C_{2}=0$). The vector $\bm{\omega}$ is kept the same as in Fig.~\ref{fig:initial-states}, while $\bm{g}$ is chosen almost perpendicular to $\bm{\omega}$. The form of $g(t)$ is the same as previously, with $g_{0}=2\omega$. The instants at which $g(t)=\omega$ (hence $C_{1}=0$) are indicated with vertical lines $t_{\pm}$.}
    \label{fig:both-sides-border}
\end{figure}
\clearpage
\end{widetext}

\begin{widetext}

\begin{figure}[H]
    \centering
    \begin{tabular}{c c c}
        & \makebox[0.4375\textwidth][c]{$\lambda=+1$}
            \makebox[0.4375\textwidth][c]{$\lambda=-1$}
        & \\
        & \includegraphics[valign=c,width=0.875\textwidth]{left_g.pdf} & \\
        \rotatebox[origin=c]{90}{$\Theta=\frac{\pi}{2}-10^{-3}$} &
        \includegraphics[valign=c,width=0.875\textwidth]{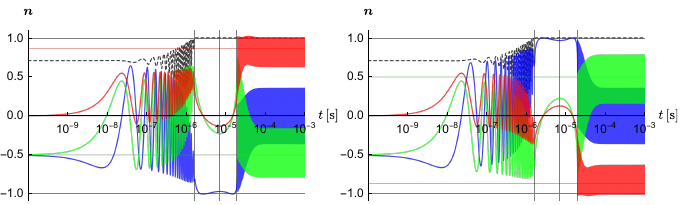} & \\
        \rotatebox[origin=c]{90}{$\bm{\Theta=\frac{\pi}{2}}$} &
        \includegraphics[valign=c,width=0.875\textwidth]{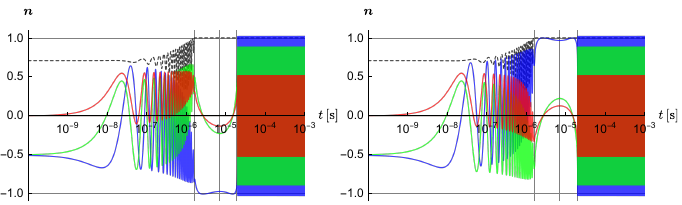} & \\
        \rotatebox[origin=c]{90}{$\Theta=\frac{\pi}{2}+10^{-3}$} &
        \includegraphics[valign=c,width=0.875\textwidth]{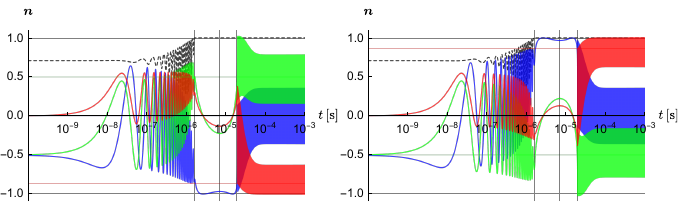} & \\
        & \includegraphics[valign=c,width=0.656\textwidth]{general_legend.pdf} & \\
    \end{tabular}
    \caption{State dynamics in proximity of the $\Theta = \pi/2$ borderline ($C_{2}=0$). The vector $\bm{\omega}$ is kept the same as in Fig.~\ref{fig:initial-states}, while $\bm{g}$ is almost perpendicular to $\bm{\omega}$. The form of $g(t)$ is the same as previously, with $g_{0}=2\omega$. The instants at which $g(t)=\omega$ (hence $C_{1}=0$) are indicated with vertical lines.}
    \label{fig:near-border}
\end{figure}
\end{widetext}

At the angle $\Theta=\pi/2$, the evolution equations exhibit a structural instability.\\
Figure~\ref{fig:near-border} presents the state dynamics in the proximity of this critical point.

In the very close vicinity of the $\Theta=\pi/2$ borderline (for the parameters chosen here, in the range about $\frac{\pi}{2}\pm10^{-3}$), the final state is no longer the eigenstate of the observable. However, until the angle $\Theta$ becomes even closer to $\frac{\pi}{2}$, the final state remains close to the eigenstate of the observable in terms of the mean value and the associated probabilities. Strictly for $\Theta=\frac{\pi}{2}$, the state oscillates furiously without any preference for one or another eigenstate of the observable. 

\subsection{Very small \texorpdfstring{$g_0$}{g0} compared with \texorpdfstring{$\omega$}{omega}}

When the $g$-potential has a sufficiently low value with a limited width, oscillations may persist, and the final state will be the eigenstate of the observable only approximately.

\begin{widetext}

\begin{figure}[H]
    \centering
    \begin{tabular}{c c c}
        & \makebox[0.4375\textwidth][c]{$\lambda=+1$}
            \makebox[0.4375\textwidth][c]{$\lambda=-1$}
        & \\
        & \includegraphics[valign=c,width=0.875\textwidth]{weak_g.pdf} & \\
        \rotatebox[origin=c]{90}{$g_0=\omega/10^{2.5}$} &
        \includegraphics[valign=c,width=0.875\textwidth]{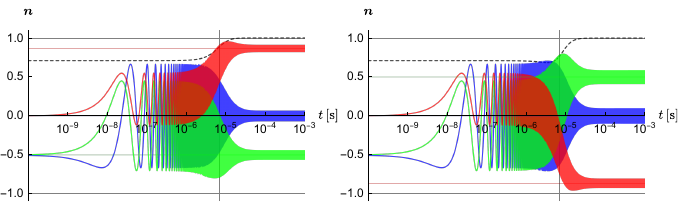} & \\
        & \includegraphics[valign=c,width=0.656\textwidth]{general_legend.pdf} & \\
    \end{tabular}
    \caption{State dynamics for an even smaller $g$-term scale, $g_{0}=10^{5.5}\hbar/\mathrm{s}$, for which the evolution departs from the measurement regime considered above. The same $\bm{\omega}$, $\bm{g}$, and functional form of $g(t)$ as in Fig.~\ref{fig:initial-states} are used. $\bm{n}_{0} = [0,-\frac{1}{2},-\frac{1}{2}]$.}
    \label{fig:small-scale}
\end{figure}
\end{widetext}

In this case, precession survives, and the final state is the approximate eigenstate of the observable only in terms of its mean value and associated probabilities.

\section{Local quasilinear measurement on entangled states}\label{sec:entangled}
Because the evolution equation \eqref{eq:measurement-evolution} is integrated to the global Kraus-like form \eqref{eq:kraus-like-solution}, local quasilinear measurements are well defined on entangled states. 

To show this, let us consider such a local action in the channel $A$ on the general state in the tensor product space $\mathcal{H}_A\otimes\mathcal{H}_B$ of two 2-dimensional quantum systems $A$ and $B$, namely on

\pagebreak

\begin{equation}
\begin{split}
&\rho_{0}=\\
&=\frac{1}{4}\left(I\otimes I+\bm{n}_{\mathrm{A}0}\cdot\bm{\sigma}\otimes I
+I\otimes \bm{n}_{\mathrm{B}0}\cdot\bm{\sigma}
+T_{ij}\,\sigma_{i}\otimes\sigma_{j}\right).
\end{split}
\label{eq:twoqubit-state}
\end{equation}

Using \eqref{eq:kraus-like-solution} and \eqref{eq:k-operator-evolution}, we obtain that 

\begin{equation}
\begin{split}
\rho_{\lambda}(t)
&=\frac{\bigl(K_{\lambda}(t)\otimes I\bigr)\rho_{0}\bigl(K_{\lambda}^{\dagger}(t)\otimes I\bigr)}
{\Tr{\bigl(K_{\lambda}(t)\otimes I\bigr)\rho_{0}\bigl(K_{\lambda}^{\dagger}(t)\otimes I\bigr)}}=\\
&=\frac{1}{4}\Biggl(
\frac{2\,K_{\lambda}(t)\rho_{\mathrm{A}0}K_{\lambda}^{\dagger}(t)}{\Tr{\bigl(K_{\lambda}(t)\rho_{\mathrm{A}0}K_{\lambda}^{\dagger}(t)\bigr)}}\otimes I+\\
&+\frac{K_{\lambda}(t)K_{\lambda}^{\dagger}(t)}{\Tr{\bigl(K_{\lambda}(t)\rho_{\mathrm{A}0}K_{\lambda}^{\dagger}(t)\bigr)}}\otimes\left(\bm{n}_{\mathrm{B}0}\cdot\bm{\sigma}\right)+\\
&+T_{ij}\,
\frac{K_{\lambda}(t)\sigma_{i}K_{\lambda}^{\dagger}(t)}{\Tr{\bigl(K_{\lambda}(t)\rho_{\mathrm{A}0}K_{\lambda}^{\dagger}(t)\bigr)}}\otimes\sigma_{j}
\Biggr).
\end{split}
\label{eq:postmeasurement-twoqubit}
\end{equation} 

\noindent The subscript $\lambda$ corresponds to two possible trajectories related to the measurement outcome in the channel $A$.

The local densities evolve according to
\begin{widetext}
\begin{subequations}\label{eq:evolved-reduced-states}
    \begin{align}
&\rho_{\mathrm{A}\lambda}(t)
=\TrB{\rho_{\lambda}(t)}
=\frac{K_{\lambda}(t)\rho_{\mathrm{A}0}K_{\lambda}^{\dagger}(t)}{\Tr{\left(K_{\lambda}(t)\rho_{\mathrm{A}0}K_{\lambda}^{\dagger}(t)\right)}},
\label{eq:evolved-A}\\
\begin{split}
&\rho_{\mathrm{B}\lambda}(t)
=\TrA{\rho_{\lambda}(t)}=\frac{1}{2}\Biggl(
I+\frac{\left(
n_{\mathrm{B}0j}\,\Tr{\left(K_{\lambda}(t)K_{\lambda}^{\dagger}(t)\right)}+T_{ij}\,\Tr{\left(K_{\lambda}(t)\sigma_{i}K_{\lambda}^{\dagger}(t)\right)}
\right)\sigma_{j}}{2\,\Tr{\left(K_{\lambda}(t)\rho_{\mathrm{A}0}K_{\lambda}^{\dagger}(t)\right)}}\Biggr).
\end{split}
\label{eq:evolved-B}
\end{align}
\end{subequations}

\end{widetext}

The standard projective measurement, by means of the von Neumann rule, leads to analogous formulas with $K_{\lambda}(t)$ replaced by the projectors $\Pi_{\lambda}$.
Below, we give an explicit example of a quasilinear measurement with an entangled state.
\pagebreak

\begin{widetext}

\begin{figure}[H]
    \centering
    \begin{tabular}{l c c}
        & \makebox[0.4375\textwidth][c]{$\lambda=+1$}
            \makebox[0.4375\textwidth][c]{$\lambda=-1$}
        & \\
        A & \includegraphics[valign=c,width=0.875\textwidth]{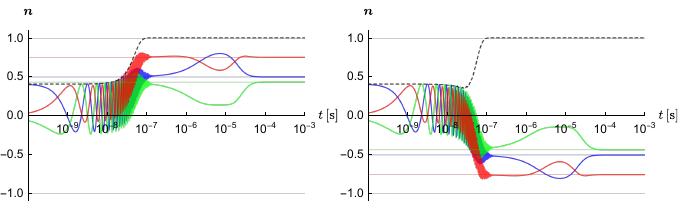} & \\
        B & \includegraphics[valign=c,width=0.875\textwidth]{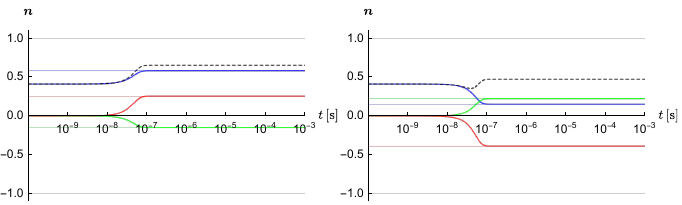} & \\
        & \includegraphics[valign=c,width=0.750\textwidth]{general_legend.pdf} & \\
        \end{tabular}
    \caption{The dynamics of subsystems $A$ and $B$ of the entangled system of two qubits subject to quasilinear measurement in the channel $A$. Vectors $\bm{\omega}$ and $\bm{g}$ (hence $\Theta$) are kept the same as in Fig.~\ref{fig:initial-states}. The initial state is chosen according to \eqref{eq:twoqubit-state} with $\bm{n}_{\mathrm{A}0}=\bm{n}_{\mathrm{B}0}=[0,0,1/\sqrt{6}]$, $T=\diag\bigl[1/\sqrt{6},-1/\sqrt{6},1/\sqrt{3}\bigr]$. The thin lines mark the final state according to the standard von Neumann projective state update rule.}
    \label{fig:entangled-local}
\end{figure}
\end{widetext}

\section{Analytic solution to the evolution equation}\label{sec:analytic}

In this section we demonstrate an analytic solution to the quasilinear state evolution equation~\eqref{eq:bloch-evolution}, or equivalently to Eqs.~\eqref{eq:kraus-like-solution} and~\eqref{eq:k-operator-evolution}, specified to the case of parallel and anti-parallel configurations of the vectors $\bm{\omega}(\alpha,\beta)$ and $\bm{g}(t,\theta,\Theta)$. This solution has an immediate link with the Stern--Gerlach experiment, where locally and close to the particle trajectories, magnetic fields are parallel.

The parallel configuration corresponds to $\Theta=0$ and $\theta=\alpha$, while the antiparallel configuration corresponds to $\theta=\alpha+\pi$ and can be obtained from the parallel configuration by the map $g(t)\rightarrow -g(t)$. Both cases can be realized by the replacement of $g(t)$ by $\lambda g(t)$, $\lambda=\pm1$. In that configuration $\bm{\omega}=\omega\hatbm{\omega}(\alpha,\beta)$ and $\bm{g}=\lambda g(t)\hatbm{\omega}(\alpha,\beta)$, with $\hatbm{\omega}(\alpha,\beta)=(\sin\alpha\cos\beta,\sin\alpha\sin\beta,\cos\alpha)$. By means of Eqs.~\eqref{eq:kraus-like-solution} and~\eqref{eq:k-operator-evolution}, adapted to the two-level system case by using Eqs.~\eqref{eq:observable-spectrum} and~\eqref{eq:driving-generator}, we obtain the following form of the operator $K(t)$ defined in Eqs.~\eqref{eq:kraus-like-solution} and~\eqref{eq:k-operator-evolution}:

\begin{equation}
\begin{split}
&K_{\lambda}(t)=\\
&=
\left(
\cosh\left(\frac{\lambda}{2}\Gamma(t)\right)I
+
\sinh\left(\frac{\lambda}{2}\Gamma(t)\right)\hatbm{\omega}(\alpha,\beta)\cdot\bm{\sigma}
\right)\times\\
&\times\left(
\cos\left(\frac{\omega t}{2\hbar}\right)I
-
i\sin\left(\frac{\omega t}{2\hbar}\right)\hatbm{\omega}(\alpha,\beta)\cdot\bm{\sigma}
\right).
\end{split}
\label{eq:k-parallel}
\end{equation}
Therefore, by means of Eq.~\eqref{eq:kraus-like-solution} we obtain the corresponding Bloch vector $\bm{n}(t)$,

\begin{widetext}

\begin{equation}
\bm{n}_{\lambda}(t)
=
\frac{
\frac{\cos\left(\frac{\omega t}{\hbar}\right)}{\cosh\left(\lambda\Gamma(t)\right)}\bm{n}_0
+
\left(
\tanh\left(\lambda\Gamma(t)\right)
+
\left(
1-\frac{\cos\left(\frac{\omega t}{\hbar}\right)}{\cosh\left(\lambda\Gamma(t)\right)}
\right)(\hatbm{\omega}\cdot\bm{n}_0)
\right)\hatbm{\omega}
+
\frac{\sin\left(\frac{\omega t}{\hbar}\right)}{\cosh\left(\lambda\Gamma(t)\right)}(\hatbm{\omega}\times\bm{n}_0)
}{
1+\tanh\left(\lambda\Gamma(t)\right)(\hatbm{\omega}\cdot\bm{n}_0)
},
\label{eq:bloch-parallel}
\end{equation}
\end{widetext}

\noindent Here $\Gamma(t)=\frac{1}{\hbar}\int_0^t g(\tau)\,d\tau$. For the inverted Morse potential $g_{\mathrm{IM}}(t)$ given by Eq.~\eqref{eq:inverse-morse-potential}, we obtain
\begin{subequations}
\begin{equation}
\Gamma_{\mathrm{IM}}(t)
=
9.3\times10^9\left(\frac{2(1-e^{-t\kappa})^2}{\kappa}\right),
\end{equation}
while for the Stern--Gerlach potential $g_{\mathrm{SG}}(t)$; see Eq.~\eqref{eq:sg-potential},
\begin{equation}
\Gamma_{\mathrm{SG}}(t)
=
\int_0^t d\tau\,
\left(
\frac{2\mu_B}{\hbar}\frac{\mu_B\beta^2}{2m_{Ag}}\,\tau^2\,\mathcal{S}(\tau)
\right).
\end{equation}
\end{subequations}
\begin{figure}[H]
\centering
\includegraphics[width=\linewidth]{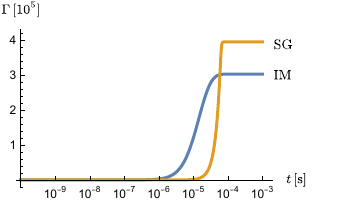}
\caption{Integrated potentials $\Gamma_{\mathrm{IM}}(t)$ and $\Gamma_{\mathrm{SG}}(t)$ discussed in the text.}
\label{fig:integrated-potential}
\end{figure}

Now, it is important to observe that in both cases, SG and IM, the integrated potential $\Gamma(t)$ demonstrated in Fig.~\ref{fig:integrated-potential} tends to a large constant value. As is evident from Fig.~\ref{fig:potentials}, this is a consequence of the positivity of the potentials $\bm{g}_{\mathrm{IM}}$ and $\bm{g}_{\mathrm{SG}}$, its fast decrease, and the extremely large value of $\hbar^{-1}$. As a result, in the Bloch-vector formula, $\cosh\left(\lambda\Gamma(t)\right)$ takes enormous values of the order of $10^{10^5}$ or larger. Thus, at $10^{-4}\,\mathrm{s}$ the Bloch vector $\bm{n}$ practically attains the form $\bm{n}(t)=\lambda\hatbm{\omega}(\alpha,\beta)$, representing the projectors $\frac{1}{2}\left(I+\lambda\hatbm{\omega}(\alpha,\beta)\cdot\bm{\sigma}\right)$ and corresponding to eigenvectors of the observable $\Omega(\bm{\omega})$; see Eq.~\eqref{eq:observable-spectrum}.

Such evolution of the state can be explained by the behavior of the operator $K(t)$ in time. Let us return to the global evolution equation~\eqref{eq:kraus-like-solution} and normalize the operator $K(t)$, taking into account that the product $K(t)K^{\dagger}(t)$ is nonnegative definite, so also its trace is nonnegative. The normalized operator is defined by
\begin{equation}
\mathcal{K}_{\lambda}(t)
=
\frac{K_{\lambda}(t)}{\sqrt{\Tr\left(K_{\lambda}(t)K_{\lambda}^{\dagger}(t)\right)}}.
\label{eq:k-normalized}
\end{equation}
From Eq.~\eqref{eq:kraus-like-solution} it follows that normalization~\eqref{eq:k-normalized} does not disturb the evolution of the state. In the considered case, by means of Eq.~\eqref{eq:k-parallel}, $\Tr\left(K_{\lambda}(t)K_{\lambda}^{\dagger}(t)\right)=2\cosh\left(\lambda\Gamma(t)\right)$, which leads to the following form of $\mathcal{K}_{\lambda}(t)$:

\begin{equation}
\begin{split}
&\mathcal{K}_{\lambda}(t)=\\
&=
\left(
\frac{\cosh\left(\frac{\lambda}{2}\Gamma(t)\right)}{\sqrt{2\cosh\left(\lambda\Gamma(t)\right)}}I
+
\frac{\sinh\left(\frac{\lambda}{2}\Gamma(t)\right)}{\sqrt{2\cosh\left(\lambda\Gamma(t)\right)}}\hatbm{\omega}(\alpha,\beta)\cdot\bm{\sigma}
\right)\times\\
&\times\left(
\cos\left(\frac{\omega t}{2\hbar}\right)I
-
i\sin\left(\frac{\omega t}{2\hbar}\right)\hatbm{\omega}(\alpha,\beta)\cdot\bm{\sigma}
\right).
\end{split}
\label{eq:k-normalized-explicit}
\end{equation}
From Fig.~\ref{fig:integrated-potential} we see that for $t > 10^{-4}\,\mathrm{s}$, $\Gamma(t)$ reaches large asymptotic values and, in consequence, the coefficients in the first factor in Eq.~\eqref{eq:k-normalized-explicit}, in front of the unit matrix and $\hatbm{\omega}\cdot\bm{\sigma}$, attain the values $\frac{1}{2}$ and $\frac{\lambda}{2}$, respectively. Calculating the product of the two terms in Eq.~\eqref{eq:k-normalized-explicit}, we finally obtain that for $t > 10^{-4}\,\mathrm{s}$ the operator $\mathcal{K}(t)$ takes the form
\begin{equation}
\mathcal{K}_{\lambda}(t)
=
\frac{1}{2}\left(I+\lambda\hatbm{\omega}(\alpha,\beta)\cdot\bm{\sigma}\right)
e^{-i\lambda\frac{\omega t}{2\hbar}}
=
\Pi_{\lambda}(\bm{\omega})\,e^{-i\lambda\frac{\omega t}{2\hbar}},
\label{eq:k-projector-limit}
\end{equation}
where, according to Eq.~\eqref{eq:observable-spectrum}, $\Pi_{\lambda}(\bm{\omega})$ is the projector from the spectral decomposition of the observable $\Omega(\bm{\omega})$. Thus, Eq.~\eqref{eq:kraus-like-solution} evolves in a very short time from
\begin{equation*}
\rho(t)
=
\frac{K(t)\rho_0 K^{\dagger}(t)}{\Tr\left(K(t)\rho_0 K^{\dagger}(t)\right)}
\end{equation*}
to
\begin{equation}
\rho_{\mathrm{out}}
=
\frac{\Pi_{\lambda}(\bm{\omega})\rho_0\Pi_{\lambda}(\bm{\omega})}
{\Tr\left(\Pi_{\lambda}(\bm{\omega})\rho_0\Pi_{\lambda}(\bm{\omega})\right)},
\label{eq:projection-limit}
\end{equation}
i.e., it finally reproduces the von Neumann projection practically perfectly.

\section{The Stern--Gerlach experiment}\label{sec:sg}

\begin{figure}[H]
\centering
\includegraphics[width=\linewidth]{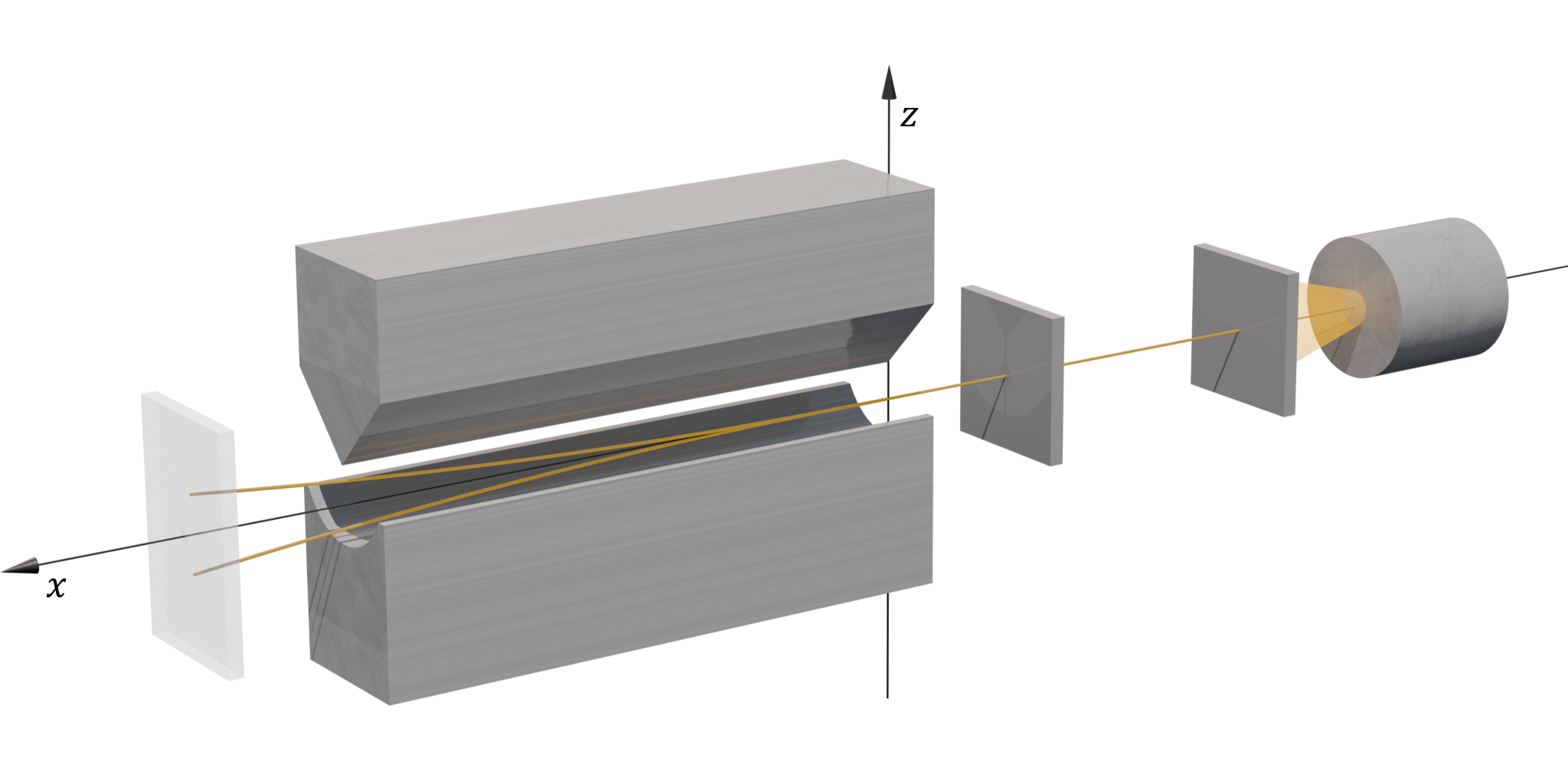}
\caption{Idealized Stern--Gerlach arrangement considered in the text.}
\label{fig:sg-setup}
\end{figure}

Now, we return to the measurement in the famous experiment by Stern and Gerlach \cite{Potel2005SG,Alstrom1982SG,Platt1992SG,Wennerstrom2012SG,Sahoo2023AnnPhys}. Because our analysis is restricted to quantum systems with a finite number of degrees of freedom, the atomic coordinates---and thus the translational motion of the atoms---are treated classically. We consider the original Stern--Gerlach arrangement employing a beam of silver atoms. The experimental setup is shown in Fig.~\ref{fig:sg-setup}.

The two magnets generate a static magnetic field $\bm{B}$, which, in the region near the particle beam, can be approximated as a superposition of dipole $\bm{b}$ and quadrupole $\bm{q}$ components. We assume an idealized scenario in which all atoms have identical initial velocity vectors $\bm{V}$ directed along the $x$-axis. The trajectories of atoms propagating through the inhomogeneous magnetic field can be determined with sufficient accuracy using semi-classical methods, for example, by applying the Ehrenfest theorem. The time dependence of the magnetic field $\bm{q}$ results from its inhomogeneity and the particle flux motion. According to the SG device configuration, we assume that the fields $\bm{b}$ and $\bm{q}$ are perpendicular to the $x$-axis and that $\bm{b}$ is oriented along the $z$-axis. The Maxwell equations $\bm{\nabla}\cdot\bm{B}=0$ and $\bm{\nabla}\times\bm{B}=0$ imply that both components of $\bm{q}$ are nonzero; however, as follows from semi-classical calculations, the $y$-axis contribution is negligible. Therefore, we accept the approximation that this contribution is zero. 

According to the physical content of the Stern--Gerlach experiment, the quantum observable $\Omega$ will be defined as a rescaled projection of the one-half spin dipole magnetic moment $\bm{\mu}=\gamma\bm{S}=\gamma\frac{\hbar}{2}\bm{\sigma}$ in a constant magnetic field $\bm{b}$,
\begin{equation}
\Omega(\bm{b}) = \bm{\mu}\cdot\bm{b} = \hbar\gamma\,\bm{b}\cdot\frac{1}{2}\bm{\sigma},
\label{eq:sg-observable-def}
\end{equation}
i.e.\ $\bm{\omega}=\hbar\gamma\bm{b}$ and $\Omega$ is identical with the classical Hamiltonian of a magnetic dipole in the magnetic field. Here $\gamma = \mathrm{g}Q/2m$ is the gyromagnetic ratio with the Land\'e factor $\mathrm{g}$, the charge of the particle $Q$ and its mass $m$. Atoms of silver utilized in Stern--Gerlach experiment have a single unpaired electron on the orbital $S$, so $\mathrm{g}=2$, $Q=e$ (electron charge), and $m=m_e$, hence $\gamma = \gamma_e = e/m_e = 2\mu_B/\hbar = 1.758736 \times 10^{11}/(\mathrm{s}\,\mathrm{T})$. Here the Bohr magneton $\mu_B = 5.788 \times 10^{-5}\,\mathrm{eV}/\mathrm{T}$ and $\hbar = 6.582 \times 10^{-16}\,\mathrm{eV}\cdot\mathrm{s}$. Therefore, operator $\Omega$ from \eqref{eq:sg-observable-def} takes the form.
\begin{equation}
\Omega(\bm{b}) = 2\mu_B\,\bm{b}\cdot\frac{1}{2}\bm{\sigma}
\equiv
\bm{\omega}\cdot\frac{1}{2}\bm{\sigma}.
\label{eq:sg-observable-omega}
\end{equation}
so $\bm{\omega}=2\mu_B\bm{b}$. The driving operator $G(\bm{q})$, encoding device configurations used in the measurement, is defined by its identification with the scalar product of the particle dipole magnetic moment $\bm{\mu}$ and the external quadrupole magnetic field $\bm{q}$ responsible for a field gradient. Therefore,

\begin{equation}
G(\bm{q}) = 2\mu_B\,\bm{q}\cdot\frac{1}{2}\bm{\sigma}
\equiv
\bm{g}\cdot\frac{1}{2}\bm{\sigma}.
\label{eq:sg-driving-operator}
\end{equation}
so $\bm{g}=2\mu_B\bm{q}$. Finally, in this case Eq.~\eqref{eq:bloch-evolution} takes the form
\begin{equation}
\frac{d\,\bm{n}(t)}{dt}
=
\gamma\left(
\bm{b}\times\bm{n}(t)
+
\lambda\,\bm{q}
-
\lambda\bm{n}(t)\left(\bm{q}\cdot\bm{n}(t)\right)
\right).
\label{eq:sg-bloch-evolution}
\end{equation}
The time dependence of the field $\bm{q}(t)=q(t)\hatbm{q}$, given in the atom co-moving frame and calculated from classical mechanics, leads to the approximate magnetic potential $q(t)$ along the particle trajectory, namely
\begin{equation}
g_{\mathrm{SG}}(t)
=
2\mu_B\,q(t)
=
2\mu_B\left(\frac{\mu_B\beta^2}{2m_{Ag}}\right)t^2\,\mathcal{S}(t),
\label{eq:sg-potential-derivation}
\end{equation}
with the smooth switch-off function $\mathcal{S}(t)$ already mentioned in the sec.~\ref{sec:numerical-examples}.

According to the standard convention, we introduce the distance parameter $L=V t$. This leads to the effective form of the magnetic field $\bm{q}(L)$ along the particle trajectory:
\begin{equation}
\bm{q}(L)
=
\left(\frac{\mu_B\beta^2}{2m_{Ag}V^2}\right)L^2\mathcal{S}(L)\hatbm{z},
\label{eq:quadrupole-field-distance}
\end{equation}
where $\mathcal{S}(L)$ is a $\mathcal{S}(t)$ function with $t=L/V$ with $L_\mathrm{end}=0.035\,m$ and $L_{\mathrm{w}}=0.001\,m$. We also specify
$b=0.1\,\mathrm{T}$, $\beta=10^3\,\mathrm{T/m}$, and
$V=550\,\mathrm{m/s}$. As follows from Eq.~\eqref{eq:quadrupole-field-distance}, the force action on atoms in the beam direction ($x$-axis) also vanishes, so atoms move in this direction with a constant velocity $V$ while in the $z$-direction the particle is accelerated.

Having the form~\eqref{eq:quadrupole-field-distance} of the magnetic field, we can eventually find the Bloch-vector evolution from the equation

\begin{equation}
\begin{split}
&\frac{d\bm{n}(L)}{dL}
=\\
&=
\frac{\gamma}{V}
\left(
\bm{b}\times\bm{n}(L)
+
\lambda
\bm{q}(L)-\lambda\bm{n}(L)\left(\bm{q}(L)\cdot\bm{n}(L)\right)
\right),
\end{split}
\label{eq:bloch-distance}
\end{equation}
being Eq.~\eqref{eq:sg-bloch-evolution} written in terms of the distance $L$. Notice, however, that in the adopted approximation the fields $\bm{b}$ and $\bm{q}$ are parallel. Therefore the solution must be of the form~\eqref{eq:bloch-parallel}, with $t$ replaced by $L/V$, $q$ oriented as in Eq.~\eqref{eq:quadrupole-field-distance}, and $\bm{b}=b\hatbm{\omega}$, where $\hatbm{\omega}=[0,0,1]$, $\hatbm{\omega}\cdot\bm{n}_0=n_{03}$, $\hatbm{\omega}\times\bm{n}_0=[-n_{02},n_{01},0]$, $\bm{n}_0=[n_{01},n_{02},n_{03}]$, and

\pagebreak

\begin{equation}
\Gamma_{\mathrm{SG}}(L)
=
\int_0^L dl\,
\left(
\frac{2\mu_B}{\hbar}\frac{\mu_B\beta^2}{2m_{Ag}}\frac{l^2}{V^3}\mathcal{S}(l)
\right).    
\end{equation}

Then Eq.~\eqref{eq:bloch-parallel} takes the form
\begin{widetext}
\begin{equation}
\bm{n}_{\lambda}(L)
=
\frac{
\frac{\cos\left(\frac{2\mu_B b L}{\hbar V}\right)}{\cosh\left(\lambda\Gamma(L)\right)}\bm{n}_0
+
\left(
\tanh\left(\lambda\Gamma(L)\right)
+
\left(
1-\frac{\cos\left(\frac{2\mu_B b L}{\hbar V}\right)}{\cosh\left(\lambda\Gamma(L)\right)}
\right)n_{03}
\right)\hatbm{\omega}
+
\frac{\sin\left(\frac{2\mu_B b L}{\hbar V}\right)}{\cosh\left(\lambda\Gamma(L)\right)}(\hatbm{\omega}\times\bm{n}_0)
}{
1+\tanh\left(\lambda\Gamma(L)\right)n_{03}
}.
\label{eq:sg-bloch-solution}
\end{equation}

\end{widetext}
The following figure presents the evolution of the spin state of a silver atom in the Stern--Gerlach experiment, according to quasilinear measurement dynamics. As is well known, the atoms in the original Stern--Gerlach experiment are initially in a fully depolarized state.

\begin{widetext}

\begin{figure}[H]
    \centering
    \begin{tabular}{c c c}
        & \makebox[0.4375\textwidth][c]{$\lambda=+1$}
            \makebox[0.4375\textwidth][c]{$\lambda=-1$}
        & \\
        \rotatebox[origin=c]{90}{$\bm{n}_0=\bm{0}$} &
        \includegraphics[valign=c,width=0.875\textwidth]{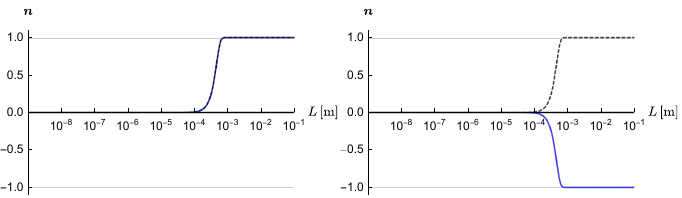} & \\
        & \includegraphics[valign=c,width=0.4375\textwidth]{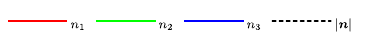} & \\
        \end{tabular}
    \caption{State evolution in the Stern--Gerlach experiment according to the quasilinear measurement dynamics. Note that $n_1(L)\equiv 0$ and $n_2(L)\equiv 0$.}
    \label{fig:sg-evolution}
\end{figure}

\end{widetext}

In this case, where the initial state is perfectly symmetric, and the geometry of the measurement setup is kept as simple as possible, we obtain a very straightforward picture. The transformation consists solely of increasing (or decreasing) the value of the third component of the state vector from $0$ to $+1$ or $-1$, which corresponds to $\pm\frac12$ spin projection on the $z$-axis. With the parameters chosen here, the state begins to change immediately upon entering the magnet slit, with an exponentially growing rate, and completes this transition within less than $1\,\mathrm{mm}$. At this point, the atom's magnetic moment is already aligned parallel or antiparallel to the $z$-axis. The choice between the two orientations is decided by the value of the $\lambda$ parameter. Since the other components remain zero, the norm of the state vector $|\bm{n}(L)|$ increases along with the absolute value of its third component. After the state transformation is complete, the norm is equal to $1$, so the state is pure---in fact, one of the two eigenstates of $\Omega$. The atom's subsequent passage through the magnet's slit in the presence of the field gradient serves only to separate streams of atoms with two defined orientations of their magnetic moment along the $z$-axis. Although the atoms in a standard SG experiment are initially in a fully depolarized state, it is instructive to consider the case when they are prepared in a specific polarized state before entering the SG apparatus.

\begin{widetext}

\begin{figure}[H]
    \centering
    \begin{tabular}{c c c}
        & \makebox[0.4375\textwidth][c]{$\lambda=+1$}
            \makebox[0.4375\textwidth][c]{$\lambda=-1$}
        & \\
        \rotatebox[origin=c]{90}{$\bm{n}_0=\frac{1}{\sqrt{2}}\left[\frac{1}{4},\frac{\sqrt{3}}{4},\frac{\sqrt{3}}{2}\right]$} &
        \includegraphics[valign=c,width=0.875\textwidth]{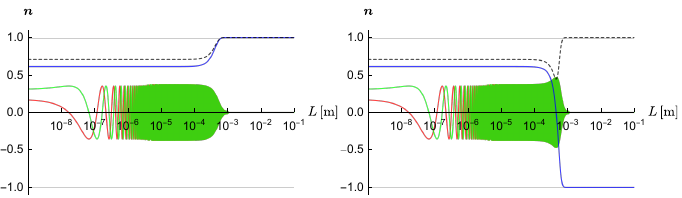} & \\
        & \includegraphics[valign=c,width=0.4375\textwidth]{legend_short.pdf} & \\
        \end{tabular}
    \caption{The evolution of the initially polarized state in the Stern--Gerlach system according to quasilinear dynamics.}
    \label{fig:sg-polarized}
\end{figure}
\end{widetext}

This case differs from the previous one in that the initial state had nonzero components in the direction perpendicular to $\hatbm{\omega}=\hatbm{z}$. Here, the state vector undergoes precession around the $z$-axis until it aligns either parallel or antiparallel to the $ z$-axis, as in the case of a depolarized state. During this process, the state's norm increases to $1$. If the state was initially partially polarized in the opposite direction to the orientation it is to attain, for a particular choice of $\lambda$, the state undergoes transient depolarisation and eventually repolarises to a pure state.

It is worth noting the rate at which this process occurs. The following figure presents the rate of change $|\bm{n}'(L)|$ of the state vector starting from the two previously mentioned initial states.

\begin{widetext}

\begin{figure}[H]
    \centering
    \begin{tabular}{c c c}
        & \makebox[0.4375\textwidth][c]{$\lambda=+1$}
            \makebox[0.4375\textwidth][c]{$\lambda=-1$}
        & \\
        \rotatebox[origin=c]{90}{$\bm{n}_0=\bm{0}$} &
        \includegraphics[valign=c,width=0.875\textwidth]{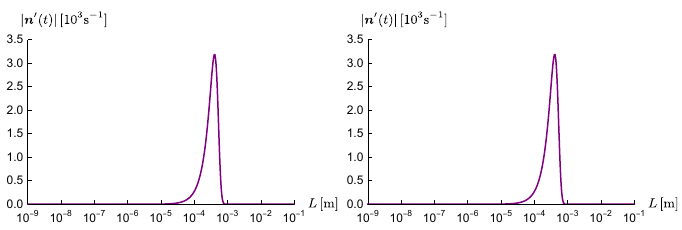} & \\
        \rotatebox[origin=c]{90}{$\bm{n}_0=\frac{1}{\sqrt{2}}\left[\frac{1}{4},\frac{\sqrt{3}}{4},\frac{\sqrt{3}}{2}\right]$} &
        \includegraphics[valign=c,width=0.875\textwidth]{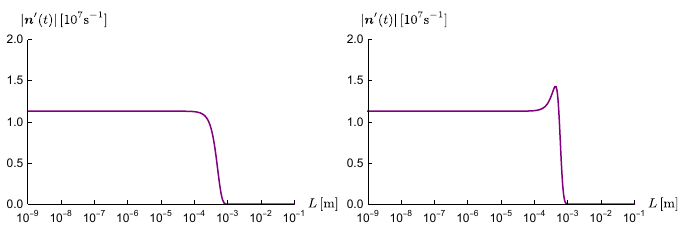} & \\
    \end{tabular}
    \caption{The rate of change of the state vector starting from the two previously mentioned initial states in the SG scenario.}
    \label{fig:sg-rate-change}
\end{figure}
\clearpage
\end{widetext}

In the case of an initially fully depolarized state, the rate of change of the state vector increases exponentially from zero until the vector approaches the eigenstate of the observable. Importantly, for the parameters chosen, this process is completed in a very short initial segment of the atom's path through the magnet's slit, when the $g$ potential is still very small compared to $\omega$. Afterward, the rate of change drops asymptotically, but very rapidly, to zero. In the case of a partially mixed state, the initial non-zero rate of change corresponds to the precession of the state vector under the rule of the $\Omega$ Hamiltonian. The rate of this precession is many times higher than the rate of change of the state vector in the previous case. Superimposed on this picture is a change similar to that observed in the depolarized state, although this time, the rate of change toward individual eigenstates is not equal. In any case, eventually, as the state approaches an eigenstate of the observable, the rate of change of the state vector decreases and finally decays to zero.

Now, the Casimir invariants $C_1$ and $C_2$ defined in Eq.~\eqref{eq:casimir-invariants} take the following form:
\begin{equation}
\begin{split}
C_{1} &= \mu_B^2\left(b^2-q(L)^2\right),\\
C_{2} &= 2\lambda b\left(\frac{\mu_B^3\beta^2}{m_{Ag}V^2}\right)L^2\mathcal{S}(L)\cos{\Theta}.    
\end{split}
\label{eq:sg-casimirs}
\end{equation}
In the Stern--Gerlach configuration considered here, both Casimir invariants remain nonzero.

Summarising, Eq.~\eqref{eq:bloch-distance} provides a theoretical description of selective measurement of a spin-$\frac12$ quantum state in terms of the postulated quasilinear evolution given in Eqs.~\eqref{eq:measurement-evolution} and~\eqref{eq:bloch-evolution}, with the Stern--Gerlach apparatus chosen as the measuring device. The effective magnetic field $\bm{q}(L)$ in Eq.~\eqref{eq:quadrupole-field-distance}, required to construct the steering term in Eq.~\eqref{eq:sg-bloch-evolution}, is determined through a classical analysis of the atomic equations of motion in the combined field $\bm{b}+\bm{q}$. It is worth noting that the Stern--Gerlach experiment can be viewed, at the classical level, as involving an overlap of two distinct processes: the transport of atoms in a magnetic field and their localization (i.e., the appearance of a spot on a photographic plate, corresponding to a detection event).
From the von Neumann point of view, after detection (measurement), the coordinate of the spot allows us to infer that, at the moment of detection, the atom state was projected onto the spin state described by $\bm{\pi}_\lambda$, where  $\lambda=\pm 1$  (post-selection). If the initial state is known, the probability of obtaining the outcome can be calculated using the Born rule.
From our perspective, the act of quantum measurement is realized as a continuous, quasilinear evolution that transforms the initial state into the final state $\bm{\pi}_\lambda$, with probabilities given by the Born rule. Analogously to the standard von Neumann measurement theory, the detection event completes the measurement by selecting a particular value of $\lambda$, thereby enabling the reconstruction of the corresponding state evolution (post-selection). This interpretation provides a complete account of the state history while preserving the conceptual framework and general features of selective measurement. Moreover, it appears that our proposal may be subject to experimental verification.

\section{Conclusions}\label{sec:conclusions}

In this article, we introduced a new form of the deterministic component of selective measurement, replacing the von Neumann projection postulate with a quasilinear evolution governed by a nonlinear generalization of the von Neumann equation (see Eq.~\eqref{eq:measurement-evolution}). In the first part of the paper, we proved that this evolution equation satisfies the quasilinearity condition, which guarantees the preservation of ensemble equivalence. As a consequence, the quasilinear evolution obeys the no-signaling principle and is therefore consistent with both quantum mechanics and Einsteinian causality.

We also derived a global, Kraus-like formal solution of this equation; see Eqs.~\eqref{eq:kraus-like-solution} and~\eqref{eq:k-operator-evolution}. Furthermore, we showed that the condition $\Tr\rho=1$, ensuring probability conservation, remains stable under the evolution. In addition, the purity of states is preserved, meaning that pure states evolve into pure states.

The evolution equation is generated by two Hermitian operators: $\Omega$, which governs the linear part and is identified with the measured observable, and $G$, which governs the nonlinear part and encodes the configuration of the experimental apparatus in relation to the quantum system under measurement. This approach eliminates the need for instantaneous state reduction without introducing additional concepts such as the quantum state of the apparatus. Importantly, the stochastic behavior of selective measurement and the Born rule remain unchanged within the quasilinear framework.

In the second part of the paper, we presented several numerical solutions to the evolution equation for quasilinear selective measurements in the case of two-level quantum systems. In this setting, the quasilinear von Neumann equation forms a system of Riccati-type differential equations; see Eq.~\eqref{eq:bloch-evolution}. We analyzed the evolution using the Bloch-vector representation of the density matrix.

Graphical results of the numerical solutions were presented and compared with the von Neumann projection. The agreement between the fundamental properties of both types of selective measurement was demonstrated, and the dynamical characteristics of the quasilinear evolution were analyzed. Practically perfect agreement of the final results of the quasilinear evolution relative to the von Neumann projection was observed for nearly all configurations of the measurement device.

We also investigated phenomena related to the structural instability of the evolution equation. We identified critical, very narrow regions in the space of device configurations where the evolution outcome deviates from the von Neumann projection. These regions could potentially be explored experimentally to test our approach.

Using the global Kraus-like representation, we also showed that local quasilinear measurements are well defined for composite systems, including entangled states. A measurement on one subsystem is represented by the operator $K_{\lambda}(t)\otimes I$, leading to conditional reduced states of both subsystems in direct analogy with the standard projective update, with $K_{\lambda}(t)$ replaced by $\Pi_{\lambda}$. This demonstrates that the quasilinear formalism provides a consistent local description of measurements in composite quantum systems.

Finally, we derived a specific analytic solution of the quasilinear evolution equation for the parallel (or antiparallel) configuration of the vectors $\bm{\omega}$ and $\bm{g}$, corresponding to the Stern--Gerlach experiment. Using this solution, we discussed the Stern--Gerlach scenario in the quasilinear context for various initial states. We also examined the dynamics of the spin state of moving silver atoms.

In this article, we have presented a detailed analysis of the dynamics of quasilinear measurement in a two-level system, in particular for spin-$\frac12$ particles. A generalization to higher-spin systems is feasible
. A quasilinear description of the Stern--Gerlach experiment for arbitrary spin particles is under preparation.

Note that the quasilinear von Neumann equation employed in this paper is the simplest, but not the only, case of the generalized quasilinear master equation \eqref{eq:qgksl-general}. Allowing non-zero Lindblad operators would yield even richer dynamics and enable combining the approach presented in this work with the effects described with the use of the GKSL master equation.

The quasilinear measurement is a proposal that is testable experimentally. A natural strategy would be to sample the evolving state at successive points along the trajectory. This could be done by extracting subensembles of particles (being the carriers of the state) at specific points, performing measurements of one or more observables on them, collecting statistics, and reconstructing the state at that point. The reconstructed state can then be directly compared with the state predicted by the quasilinear evolution model. 

Particularly interesting would be the experimental identification of critical points that occur in certain configurations of quasilinear measurements. However, in the case of the Stern--Gerlach experiment, presented in Fig.~\ref{fig:sg-evolution}, the quasilinear evolution is rather simple and unlikely to yield sufficiently distinctive signatures. Therefore, a more sophisticated setup would most likely be required.

\section*{Acknowledgments}
We thank Pawe{\l} Horodecki for the possibility of presenting some ideas and results of our work at the main session of the XVI Symposium KCIK-ICTQT Quantum Information, 7--10 May 2025, Gda{\'n}sk--Sopot, Poland. 
We gratefully acknowledge Polish high-performance computing infrastructure PLGrid (HPC Center: ACK Cyfronet AGH) for providing computer facilities and support within computational grant no. PLG/2024/017819

\section*{Author contribution}
J.R.: Conceptualization, Formal analysis, Writing (all stages). K.Ł.: Formal analysis, Visualization, Writing (all stages).

\printbibliography

\end{document}